\documentclass[10pt,final,journal]{IEEEtran}
\usepackage{amsmath,latexsym,amssymb,amsthm,array,bbm}
\usepackage{revsymb}
\usepackage{graphicx,wrapfig,color,verbatim,mathtools}
\usepackage[hmargin=3cm,vmargin=3.5cm]{geometry}

\newtheorem{theorem}{Theorem}
\newtheorem{lemma}[theorem]{Lemma}
\newtheorem{corollary}[theorem]{Corollary}
\newtheorem{proposition}[theorem]{Proposition}
\newtheorem{definition}[theorem]{Definition}
\newtheorem{remark}[theorem]{Remark}

\renewenvironment{proof}{\noindent\textbf{Proof.} }{\hfill\qed}

\newlength{\blank}
\newcommand{\dl}[1]{#1}

\newcommand{\nc}{\newcommand}

\nc{\pw}{\mt{PW}}

\nc{\arbclass}{\mt{\Omega}}

\nc{\rnc}{\renewcommand}

\nc{\mc}[1]{{\mathcal #1}}

\def\mt{\mathrm}

\nc{\lbar}[1]{\overline{#1}}
\nc{\bra}[1]{\langle#1|}
\nc{\ket}[1]{|#1\rangle}
\nc{\ketbra}[2]{|#1\rangle\!\langle#2|}
\nc{\braket}[2]{\langle#1|#2\rangle}
\nc{\proj}[1]{| #1\rangle\!\langle #1 |}
\nc{\avg}[1]{\langle#1\rangle}
\nc{\rank}{\operatorname{rank}\,}
\nc{\smfrac}[2]{\mbox{$\frac{#1}{#2}$}}
\nc{\tr}{\operatorname{Tr}}
\nc{\ox}{\otimes}

\nc{\catchset}{T}

\nc{\dg}{\dagger}
\nc{\dn}{\downarrow}

\nc{\1}{{\mathbbm{1}}}
\nc{\id}{{\operatorname{id}}}

\nc{\oxk}[1]{k_0^{#1}} 
\nc{\xk}[1]{K_0^{#1}} 
\nc{\ozec}[1]{c_0^{#1}} 
\nc{\zec}[1]{C_0^{#1}} 

\nc{\CCs}[2]{\mathbf{C}(#1\!\rightarrow\!#2)}
\nc{\aBCC}[5]{#1(#2\!\rightarrow\!#3,\!#4\!\rightarrow\!#5)}
\nc{\XBs}[4]{\mt{\Omega}(#1\!\rightarrow\!#2, #3\!\rightarrow\!#4)}
\nc{\LCs}[4]{\NC(#1\!\rightarrow\!#2, #3\!\rightarrow\!#4)}

\def\NC{\mt{NC}}
\def\SR{\mt{SR}}
\def\SE{\mt{SE}}
\def\NS{\mt{NS}}

\def\hg{H}

\begin{document}

\title{Zero-error channel capacity and simulation\protect\\ assisted by non-local correlations}

\author{Toby~S.~Cubitt, Debbie~Leung, William~Matthews and Andreas~Winter
  \thanks{Toby Cubitt is at the University of Bristol. William Matthews (corresponding author: will@northala.net) and Debbie Leung are with the Institute for Quantum Computing at the University of Waterloo. Andreas Winter is at the University of Bristol and the National University of Singapore. TSC is supported by a Leverhulme early-career fellowship and the EC project ``QAP'' (contract no.~IST-2005-15848). DL was funded by CRC, CFI, ORF, CIFAR, NSERC, and QuantumWorks. WM acknowledges the support of NSERC and QuantumWorks. AW is supported by the EC, the U.K. EPSRC, the Royal Society, and a Philip Leverhulme Prize. The CQT is funded by the Singapore MoE and the NRF as part of the Research Centres of Excellence programme. We are grateful for the hospitality of the Kavli Institute for Theoretical Physics at UCSB, where a large part of this research was performed. This research was supported in part by the NSF under Grant No.~PHY05-51164.} }

\maketitle

\begin{abstract} 
  The theory of zero-error communication is re-examined in the broader setting of using one classical channel to simulate another exactly in the presence of various classes of non-signalling correlations between sender and receiver i.e. shared randomness, shared entanglement and arbitrary non-signalling correlations. When the channel being simulated is noiseless, this is zero-error coding assisted by correlations. When the resource channel is noiseless, it is the reverse problem of simulating a noisy channel exactly by a noiseless one, assisted by correlations.  In both cases, separations between the power of the different classes of assisting correlations are exhibited for finite block lengths. The most striking result here is that entanglement can assist in zero-error communication. In the large block length limit, shared randomness is shown to be just as powerful as arbitrary non-signalling correlations for exact simulation, but not for asymptotic zero-error coding. For assistance by arbitrary non-signalling correlations, linear programming formulas for the asymptotic capacity and simulation rates are derived, the former being equal (for channels with non-zero unassisted capacity) to the feedback-assisted zero-error capacity derived by Shannon. Finally, a kind of reversibility between non-signalling-assisted zero-error capacity and exact simulation is observed, mirroring the usual reverse Shannon theorem.
\end{abstract}

\section{Introduction}
Much of classical and quantum information theory is concerned with the use of one resource (a channel, an entangled state, etc.) to simulate another. Typically errors are allowed in the simulation protocol if they vanish asymptotically as the number of resources involved grows. One then asks for the asymptotic rates of resource exchange: Shannon's channel coding theorem~\cite{shannon-BIG} tells us the asymptotic rate at which we need to make use of a given discrete memoryless channel to simulate a perfect bit channel. The quantum reverse Shannon theorem~\cite{QRST} shows that a single number associated to quantum channels, the entanglement assisted classical capacity $C_E$, determines the rate at which it can simulate another when entanglement is a free resource. Since $C_E$ reduces to the Shannon capacity for classical channels, the availability of entanglement does not affect the rate at which one classical channel can simulate another, \dl{in the setting where errors which vanish in the large block length limit are tolerated.}

Since it is often unrealistic to assume that arbitrarily long block lengths can be used in encoding and decoding, an alternative, idealised, task of \emph{zero-error coding} \cite{ZEIT} has been considered since the seminal 1956 paper of Shannon \cite{shannon} and more recently in quantum information theory \cite{Med, Duan, CCH-zero, superduper}.

For a suitable definition of decoding error probability $p_e$, both asymptotic and zero-error coding theory make statements about the region of triples $(n,k,p_e)$ which can be achieved by codes which use $n$ channel uses to transmit $k$ bits (or, equivalently, one of $2^k$ symbols). The full characterisation of this achievable region is normally far from tractable. Whereas the freedom granted by demanding only that $p_e \to 0$ as $n \to \infty$ admits simplification via random coding arguments (for example) in the asymptotic theory, the zero-error theory (which studies the restriction of the region to the plane $p_e = 0$) is attractive because the problem becomes essentially combinatorial. Nevertheless, it is a source of hard mathematical problems: In Shannon's groundbreaking work on the subject \cite{shannon} he made a conjecture (on the zero-error capacity of the pentagon channel) which had to wait over twenty years before it was proven by Lov\'{a}sz~\cite{Lovasz}. Many related open problems remain \cite{ZEIT}.

In this paper we consider both zero-error coding and the ``reverse'' problem of exact simulation of noisy channels when various types of correlations between sender and receiver are freely available. This leads to various relaxations of the combinatorial problems posed by the unassisted theory, some of which have complete and general solutions.

\section{Overview}

This section introduces the central concepts and quantities dealt with in the rest of the paper (please note that an index of notations is provided as an appendix).
$\CCs{X}{Y}$ denotes the set of discrete, memoryless, classical channels (i.e. conditional probability distributions) with inputs in $X$ and outputs in $Y$, $X$ and $Y$ being finite sets. $\aBCC{\mathbf{C}}{A}{S}{B}{T}$ means the set of bipartite conditional probability distributions, with inputs in the set $A$ and outputs in $S$ for Alice, and inputs in $B$ and outputs in $T$ for Bob. We will frequently consider bipartite distributions that are non-signalling, which we will refer to as \emph{correlations}. A \emph{class} $\arbclass$ of correlations is a subset of all possible bipartite conditional probability distributions defined by some property such that the set is closed under local operations by either party, in additon to all distributions in $\Omega$ being non-signalling. We denote the subset of $\aBCC{\mathbf{C}}{A}{S}{B}{T}$ which is in the class $\arbclass$ by $\XBs{A}{S}{B}{T}$.

Here we deal with the following classes of correlations: A bipartite channel is in $\NC$ if it can be implemented by local operations alone --- there are No Correlations between the two parties at all. Correlations belong to $\SR$ if they can be obtained using (classical) Shared Randomness (and local operations); to $\SE$ (Shared Entanglement) if they can be obtained from local operations on a shared quantum state; and to $\NS$ if the correlation is Non-Signalling in both directions: That is, the marginal distribution of Alice's output is independent of Bob's input and vice versa. Each class in this list has a strictly weaker defining property than the last, so we have $\aBCC{\NC}{A}{S}{B}{T}  \subset \aBCC{\SR}{A}{S}{B}{T} \subset \aBCC{\SE}{A}{S}{B}{T} \subset \aBCC{\NS}{A}{S}{B}{T}$.

If Alice and Bob are connected by a classical channel $\mc{N} \in \CCs{X}{Y}$ and have access to any correlation in class $\arbclass$ (shared randomness, entanglement etc.) then they can \emph{exactly simulate} $\mc{M} \in \CCs{Q}{R}$ if there is a \emph{local} protocol whereby Alice takes an input $q \in Q$ and, through local operations and a single use of $\mc{N}$ and any use of $\arbclass$, Bob produces an output $r \in R$, such that the conditional probability of $r$ given $q$ is exactly $\mc{M}(r|q)$. To say that $n$ uses of $\mc{N}$ can exactly simulate $m$ uses of $\mc{M}$ means that $\mc{N}^{\otimes n}$ can exactly simulate ${\mc{M}}^{\otimes m}$.

\begin{figure}[ht]
  \includegraphics[scale=0.5]{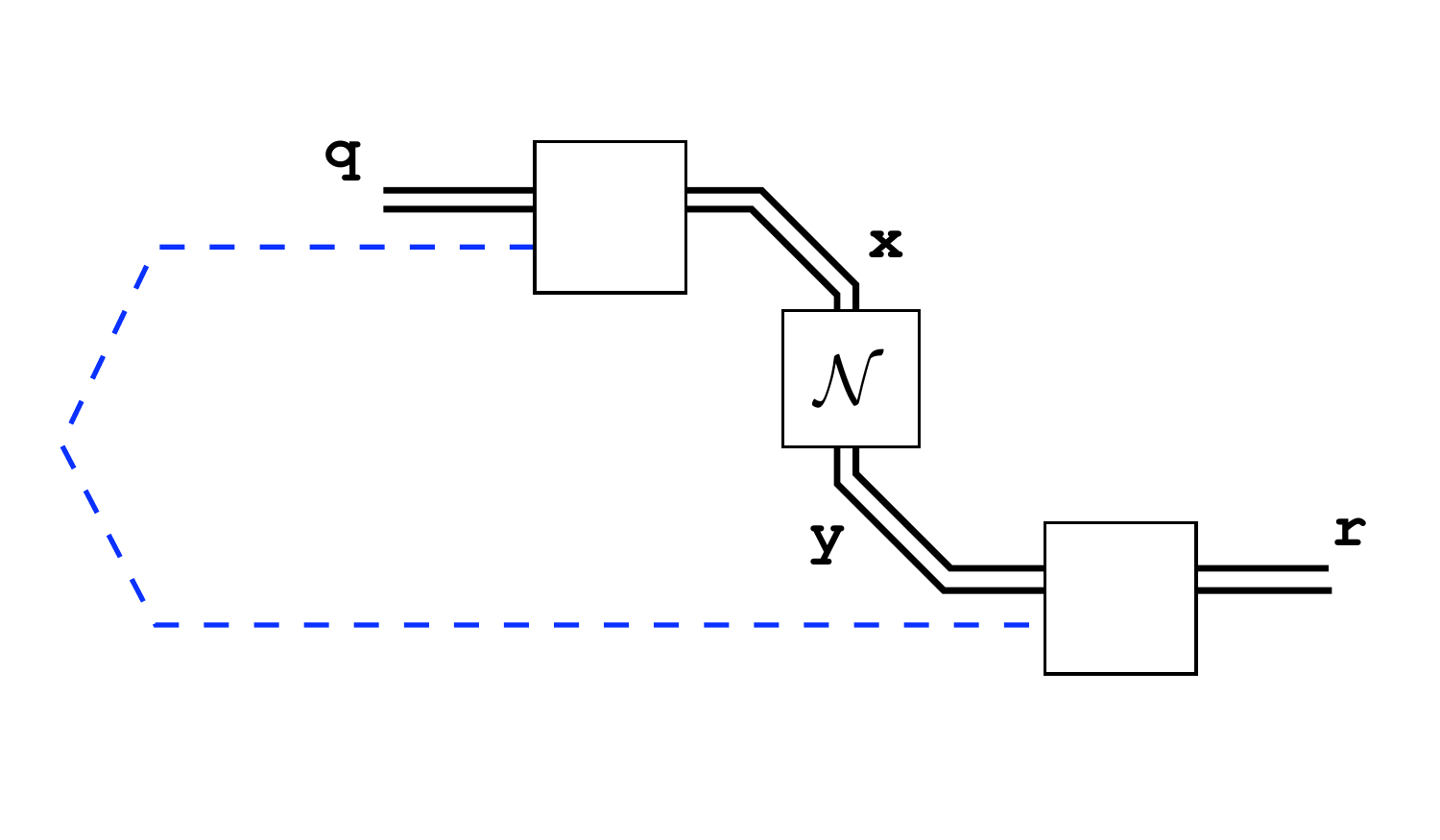}
  \centering
  \caption{Schematic representation of $\mc{M} = W[P,\mc{N}]$: A pre-shared resource (randomness, entanglement or maybe something non-physical) is portrayed by a dotted line. Alice and Bob interact with this, resulting in the non-signalling correlation $P(x,r|q,y) \in \aBCC{\mathbf{C}}{Q}{X}{Y}{R}$: Alice goes first, and obtains $x$ which she inputs into the channel $\mc{N} \in \CCs{X}{Y}$. Then, based on the channel output $y$, Bob interacts with the correlation resource, obtaining an outcome $r$. For example, if the resource is an entangled system, then w.l.o.g. both parties' interactions consist in choosing from a set of generalised measurements to perform on their local system. This all results in the channel $\mc{M} = W[P,\mc{N}] \in \CCs{Q}{R}$.}
  \label{fig:wiring}
\end{figure}

On pairs consisting of a bipartite correlation $P \in \aBCC{\arbclass}{Q}{X}{Y}{R}$ and a classical channel $\mc{N} \in \CCs{X}{Y}$ we define a bilinear map $W$ which corresponds to `wiring' Alice's output of $P$ to the input of $\mc{N}$ and the output of $\mc{N}$ to Bob's input to $P$ to produce a new classical channel $\mc{M} = W[P,\mc{N}]$, with
\begin{equation*}
  \mc{M}(r|q) := \sum_{x \in X, y \in Y} P(x,r|q,y) \mc{N}(y|x).
\end{equation*}
Because of the time ordering involved, this only makes operational sense if $\arbclass$ is non-signalling from Bob to Alice, and if it is not then $\mc{M}$ may not be a valid conditional distribution. See Figure~\ref{fig:wiring} for a diagram of the operational meaning. Set valued arguments to $W$ are given the natural interpretation as yielding the image sets of classical channels.

Since classes of correlations are closed under local operations, a channel $\mc{M} \in \CCs{Q}{R}$ can be exactly simulated by a single use of $\mc{N} \in \CCs{X}{Y}$ and correlations in $\arbclass$ if and only if
\begin{equation*}
  \mc{M} \in W[\XBs{Q}{X}{Y}{R}, \mc{N}].
\end{equation*}

Now, we can ask for the optimal use of one channel to simulate another one in the presence of some class of correlations $\arbclass$. In this paper, we shall concentrate on the simulation of perfect (i.e. identity) channels by noisy ones (``zero-error capacity'') and the reverse (``exact simulation cost'').

\begin{definition}\label{ozec}
  For a classical channel $\mc{N} \in \CCs{X}{Y}$, and free correlations from the class $\arbclass$, 
  let $\ozec{\arbclass}(\mc{N})$ denote the maximum alphabet size $c$ such that one symbol from the alphabet can be sent without error using $\arbclass$ and a single use of $\mc{N}$:
  \begin{multline*}
    \ozec{\arbclass}(\mc{N}) :=\\
	\max\{ c : \id_c \in W[\aBCC{\arbclass}{[c]}{X}{Y}{[c]}, \mc{N}]
    \},
  \end{multline*}
  where $\id_c$ is the classical identity channel on $c$ symbols.
\end{definition}
Since clearly $\ozec{\arbclass}(\mc{N}_1\ox \mc{N}_2) \geq \ozec{\arbclass}(\mc{N}_1)\ozec{\arbclass}(\mc{N}_2)$, Fekete's lemma guarantees existence of the \emph{$\arbclass$-assisted zero-error capacity} of a channel $\mc{N}$ defined by
\begin{equation*}
  \zec{\arbclass}(\mc{N})
  := \lim_{n \to \infty} \frac{1}{n} \log \ozec{\arbclass}(\mc{N}^{\ox n}).
\end{equation*}
In this paper ``$\log$'' is base 2 so this is the capacity in bits. We use ``$\ln$'' for the natural logarithm.

\begin{definition}
  For a classical channel $\mc{N} \in \CCs{X}{Y}$, and free correlations in class $\arbclass$, 
  let $\oxk{\arbclass}(\mc{N})$ denote the minimum alphabet size $k$ such that perfect transmission of one symbol of the alphabet allows exact simulation of one use of the channel:
  \begin{multline*}
    \oxk{\arbclass}(\mc{N}) :=\\
    \min \{
      k : \mc{N} \in W[\XBs{X}{[k]}{[k]}{Y}, \id_k)]
    \}.
  \end{multline*}
\end{definition}
\noindent Similarly, we define
\begin{equation*}
  \xk{\arbclass}(\mc{N})
  := \lim_{n \to \infty} \frac{1}{n} \log  \oxk{\arbclass}(\mc{N}^{\ox n})
\end{equation*}
as the asymptotic rate at which perfect classical bits must be transmitted to perfectly simulate $\mc{N}$, if correlations in class $\arbclass$ are free. The existence of the limit is once more guaranteed by Fekete's lemma, because $\oxk{\arbclass}$ is clearly submultiplicative:
\begin{equation*}
  \oxk{\arbclass}(\mc{N}_1 \ox \mc{N}_2)
  \leq \oxk{\arbclass}(\mc{N}_1) \oxk{\arbclass}(\mc{N}_2).
\end{equation*}

\subsection{Structure \dl{of the paper}}
The classical reverse Shannon theorem~\cite{CRST} assures us that in a setting of asymptotically vanishing simulation errors, all channels can reversibly simulate each other when shared randomness between sender and receiver is freely available: the rate at which $\mc{N}_1$ can simulate $\mc{N}_2$ being the ratio of their Shannon capacities, $C(\mc{N}_1)/C(\mc{N}_2)$. This remains true when entanglement and even more general non-signalling resources are shared by sender and receiver.

The exact coding and simulation problem will be shown to have a much more complex structure. In the next section we review some of the classical theory of zero-error coding, discuss the correlation assisted zero-error quantities $\ozec{\arbclass}$ and $\zec{\arbclass}$, and then show several separations between them for different classes $\arbclass$ of assisting correlation. The most striking results here are a complete solution for the non-signalling assisted case and the construction of channels where entanglement assists for the one-shot scenario (i.e. where $\ozec{\SE} > \ozec{}$).

In section~\ref{sec:sim} we explore the quantities $k_{\arbclass}$, which we show to be all different (in general) for $\arbclass \in \{ \NC, \SR, \SE, \NS \}$; and the simulation rates $\xk{\arbclass}$, which turn out to be all the same for $\arbclass \in \{ \SR, \SE, \NS \}$, and indeed are given by a simple formula. We even find a kind of combinatorial reverse Shannon theorem for zero-error communcation/noisy channel simulation in the presence of general non-signalling correlations: the simulation rate minimised over all channels with the same pattern of zeroes as the matrix $\mc{N}(y|x)$ is the same the non-signalling assisted zero-error capacity of $\mc{N}$.

We conclude with some open questions.

\section{Assisted zero-error capacities}
\label{sec:C_ass}

\subsection{Local operations and shared randomness}
We start with the least powerful resources, $\NC$ and $\SR$. The former simply describes arbitrary encoding and decoding maps. Shared randomness doesn't change anything since any value of the shared randomness will have to yield a zero-error coding if the randomised protocol does. For the same reason nothing is lost by requiring deterministic encoding and decoding maps, so the coding can be fully specified by giving a subset of input symbols to use as codewords. Thus we are in Shannon's original zero-error setting~\cite{shannon}, and we shall write $\ozec{} = \ozec{\NC} = \ozec{\SR}$ and $\zec{} = \zec{\NC} = \zec{\SR}$.

The fundamental observation is that, for zero-error coding over a channel $\mc{N} \in \CCs{X}{Y}$, two symbols can both be used as codewords only if they are not \emph{confusable}, that is, only if the corresponding output distributions have disjoint support. Therefore, a zero-error code is just a set of pairwise non-confusable input symbols in $X$, and $\ozec{}(\mc{N})$ is the largest size of such a set.

In general, it is not hard to see that for any of our resources $\arbclass$, only the pattern of zeroes in $\mc{N}(y|x)$ can affect $\ozec{\arbclass}$ and $\zec{\arbclass}$, so that the zero/one matrix $\lceil \mc{N}(y|x) \rceil$ encodes all the relevant information. This motivates the introduction of the following combinatorial representations of channels.


\begin{definition}
	The \emph{hypergraph} $\hg(\mc{N})$ of a channel $\mc{N} \in \CCs{X}{Y}$ has vertex set $X$ and hyperedges
	\[
		E(H(\mc{N})) := \{ e_{y} := \{ x : \mc{N}(y|x) > 0 \} : \forall y \in Y \} 
	\]
	capturing the equivocation of each output symbol $y \in Y$. 
\end{definition}


\dl{Note that different output symbols can give rise to the same hyperedge, so that the number of hyperedges may be less than the number of output symbols.} 

Looking back at Definition $\ref{ozec}$, let $P(\hat{z}|z,y;x)$ denote the probability distribution on Bob's output from the correlation conditional on Alice having input $z$, Bob having input $y$, and Alice having obtained output $x$. This is not well defined if $z$ never occurs for $x$, and in this case we set $P(\hat{z}|z,y;x) = 0$ (so it is not in fact a distribution). When, Bob obtains an output $y$ he knows that there is non-zero probability that Alice made input $x$ iff it belongs to the hyperedge $e_y$. The correlation $P$ yields a zero-error coding iff $\sum_{\hat{z}}P(\hat{z}|z,y;x)P(\hat{z}|z',y;x) = 0$ for all $x \in e_y$ whenever $z \neq z'$ for every hyperedge $e_y$ in $E(H(\mc{N}))$. Therefore, a correlation assisted zero-error capacities depends only on the channel hypergraph.

To compute the unassisted zero-error capacity an even coarser representation of the channel will suffice:
\begin{definition}\label{cg}
	The \emph{confusability graph} $G(\mc{N})$ of a channel $\mc{N} \in \CCs{X}{Y}$ has vertices $X$ and an edge between input symbols $x$ and $x'$ iff they are confusable, i.e.~$\sum_{y \in Y} \mc{N}(y|x)\mc{N}(y|x') > 0$.
\end{definition}

With this notation, $\ozec{}(\mc{N})$ is simply $\alpha(G(\mc{N}))$: the \emph{independence number} of $G(\mc{N})$.

	Clearly $G(\mc{N})$ can be obtained from $H(\mc{N})$ by taking the vertex set of $H$ as the vertex set of $G$ and joining vertices with an edge iff there is a hyperedge of $H$ containing both. On the other hand, given a graph $G$ with vertex set $X$, there are generally many hypergraphs on $X$ which are mapped to $G$ by this rule. Hypergraphs on a given vertex set form a lattice when ordered by inclusion of their sets of hyperedges. The supremum of the set of hypergraphs with confusability graph $G$ is the clique hypergraph of $G$, $\chi(G)$, whose hyperedges are all of the cliques in $G$. From the point of view of zero-error coding, extra hyperedges can only be a bad thing, and this represents the worst case: For all hypergraphs $H$ with a given confusability graph $G$, $\ozec{\arbclass}(H) \geq \ozec{\arbclass}(\chi(G))$.

For two graphs $G_1, G_2$ with vertex sets $X_1, X_2$ their strong product $G_1 \ox G_2$ is the graph on $X_1 \times X_2$ with an edge $\{(x_1,x_2),(z_1,z_2)\}$ iff $(\{x_1,x_2\}\in E(G_1)) \wedge (\{z_1,z_2\}\in E(G_2))$ or $(x_1 = x_2)\wedge (\{z_1,z_2\}\in E(G_2))$ or $(\{x_1,x_2\}\in E(G_1)) \wedge (z_1 = z_2)$. In terms of confusability graphs, $G(\mc{N}_1\ox\mc{N}_2) = G(\mc{N}_1)\ox G(\mc{N}_2)$. For two hypergraphs $H_i$ with vertex sets $X_i$ and edges $E_i$, ($i=1,2$), we define the product $H_1\ox H_2$ on vertex set $X_1\times X_2$ to have the hyperedges $\{e\times f : \forall\ e \in E_1, f \in E_2\}$. The hypergraph of a product channel is the product of the individual hypergraphs, and the clique hypergraph of a strong graph product is the product of the individual clique hypergraphs.

The \emph{Shannon capacity} of a graph is the asymptotic behaviour of the independence number of the strong product of $n$ copies
\[
\Theta(G) := \lim_{n\rightarrow\infty} \sqrt[n]{\alpha(G^{\ox n})}.
\]
The zero-error capacity of $\mc{N}$ is the same quantity but measured in bits per channel use
\[
	\zec{}(\mc{N}) = \log \Theta(G(\mc{N})).
\]
The smallest example where the supermultiplicitvity of $\ozec{} (= \alpha)$ is strict is the pentagon graph $C_5$, for which $\ozec{}(C_5) = 2$ but $\ozec{}(C_5^{\ox 2}) = 5$. Shannon conjectured that $\Theta(C_5) = \sqrt{5}$,
which was only shown to true by Lov\'{a}sz \cite{Lovasz}.

Determining whether $\ozec{}(\mc{N})$ is greater than a given integer $k$ is NP-complete (indeed it is trivially equivalent to $k$-CLIQUE). Whether $\Theta$ is larger than some number is not even known to be decidable.

Shannon~\cite{shannon} found an upper bound on the zero-error capacity by considering \emph{feedback} assistance. In this scenario, as soon as Bob receives an output $y$ from the channel, Alice gets to know this $y$ with perfect reliability. While this is no advantage if only a single use of the channel is made, it is sometimes useful given multiple uses (an observation Shannon attributes to Elias \cite{shannon}). Shannon goes on to give a general formula for the asymptotic feedback assisted zero-error capacity $C_{\mt{0FB}}$. It is zero whenever $\zec{}(\mc{N})$ is zero (i.e. whenever the confusability graph of the channel is complete) but otherwise is precisely the \emph{fractional packing number} of the channel hypergraph.

\begin{definition}
	\label{defi:fractional}
	A \emph{fractional packing} of a hypergraph $H$ with vertex set $V(H) = X$ is an assignment of non-negative weights $v(x) \leq 1$ to all vertices $x$ such that
  \[
    \forall e \in E(H) \quad \sum_{x \in e} v(x) \leq 1.
  \]
	A \emph{fractional covering} of a hypergraph $H$ with vertex set $V(H) = X$ is an assignment of non-negative weights $w(e) \leq 1$ to all hyperedges $e \in E(H)$ such that
	\[
	\forall\ x\in X \quad \sum_{e \ni x} w(e) \geq 1.
  	\]
  (For weights in $\{0,1\}$ we recover the combinatorial notions of packing and covering.)

The \emph{fractional packing number} $\alpha^*(H)$ is the maximum total weight allowed in fractional packing of $H$ and the \emph{fractional covering number} $\omega^*(H)$ is the minimum total weight required for a fractional covering of $H$. These are clearly dual linear programs, which for a channel hypergraph $H(\mc{N})$ have the formulation
	\begin{align*}
	    \alpha^*(H(\mc{N})) = \max \bigg\{ & \sum_{x \in X} v(x) : \forall x\in X, v(x) \geq 0,\\
	      & \sum_{y \in Y} \lceil \mc{N}(y|x) \rceil v(x) \leq 1 \bigg\} \\
	    \omega^*(H(\mc{N})) = \min \bigg\{ & \sum_{y \in Y} w(x) : \forall y\in Y, w(y) \geq 0,\\
	      & \sum_{x \in X} \lceil \mc{N}(y|x) \rceil w(y) \geq 1 \bigg\}.
	  \end{align*}
\end{definition}

Note that the possibility of redundant hyperedges in this representation (as compared with the purer one in terms of sets) has no effect on either quantity.

\dl{The fractional packing problem is always feasible.  On the other hand, the fractional covering problem is feasible if and only if the union of all hyperedges covers $X$.} Where both are feasible $\alpha^{\ast}(H) = \omega^{\ast}(H)$ by the strong duality theorem for linear programs.  
\dl{In particular, this holds for a channel hypergraph, since the fractional covering problem is always feasible (as every input symbol always results in \emph{some} output symbol occurring).}

From the definition of $\alpha^*$, we have $\alpha(G) \leq \alpha^*_{\chi}(G) := \alpha^*(\chi(G)) \leq \alpha^*(H)$ for any hypergraph with confusability graph $G$. But it turns out that $\alpha^*$ is even an upper bound on $\Theta(G)$:

\begin{proposition}
  \label{prop:duality1}
  $\alpha^*$ is multiplicative with respect to the direct hypergraph product: $\alpha^*(H_1\otimes H_2) = \alpha^*(H_1)\,\alpha^*(H_2)$.
\end{proposition}
\begin{proof}
  To show multiplicativity, the strong duality means that it suffices to show supermultiplicativity of $\alpha^*$, and submultiplicativity of $\omega^*$, i.e.
  \begin{align*}
    \alpha^*(H_1\otimes H_2)
      &\geq \alpha^*(H_1)\,\alpha^*(H_2)
\\
    \omega^*(H_1\otimes H_2)
      &\leq \omega^*(H_1)\,\omega^*(H_2).
  \end{align*}
  These are easy, because it is straightforward to confirm that the tensor product of two feasible vectors $v_1$ and $v_2$ (dual feasible vectors $w_1$ and $w_2$) for $H_1$ and $H_2$, respectively, is feasible (dual feasible) for $H_1\times H_2$.
\end{proof}

Therefore, for any integer $n$, $\alpha(G^{\ox n}) \leq \alpha^*_{\chi}(G^{\ox n}) = \alpha^*(\chi(G)^{\ox n}) = \bigl( \alpha^*_{\chi}(G) \bigr)^n$, so $\Theta(G) \leq \alpha^*_{\chi}(G)$. But this bound is often not tight. For example for the pentagon $C_5$, it yields $\Theta(C_5) \leq \frac{5}{2}$; the above two-copy consideration shows on the other hand that $\Theta(C_5) \geq \sqrt{5}$. The celebrated result of Lovasz~\cite{Lovasz} says that the lower bound is tight, $\Theta(C_5) = \sqrt{5}$. He proved this by introducing another, tighter, but still multiplicative relaxation for $\alpha(G)$, denoted $\vartheta(G)$.



\subsection{Assistance by non-signalling correlations}
Now that we have reviewed the state of the art regarding $\ozec{}$ and $\zec{}$, we go on to present our complete solution for $\ozec{\NS}$ and $\zec{\NS}$.
\begin{theorem}
  \label{NSZELP}
  For a classical channel $\mc{N} \in \CCs{X}{B}$ with hypergraph $H(\mc{N})$
  \begin{equation*}
    \ozec{\NS}(\mc{N})
    = \left\lfloor \alpha^*(H(\mc{N})) \right\rfloor
  \end{equation*}
  where $\alpha^*(H(\mc{N}))$ is the fractional packing number of $H(\mc{N})$. Being a linear program, this can be efficiently computed from the channel.

  Furthermore, since $\alpha^*$ is multiplicative, the NS-assisted zero-error capacity of a channel is
  \begin{equation*}
    \zec{\NS}(\mc{N}) = \log \alpha^*(H(\mc{N})).
  \end{equation*}
\end{theorem}
\begin{proof}
$\mc{N} \in \CCs{X}{Y}$ can exactly simulate a $g$-message identity channel with non-signalling correlations if and only if there exists $P$ in $\aBCC{\NS}{[g]}{X}{Y}{[g]}$ such that
  \begin{equation*}
    \sum_{x \in X, y \in Y} P(x,\hat{z}|z,y) \mc{N}(y|x) = \begin{cases}
      1 & \text{ if } \hat{z} = z, \\
      0 & \text{ if } \hat{z} \neq z.
	\end{cases}
  \end{equation*}
\dl{  
Without loss of generality, we can assume that a simplified form of non-signalling correlation is used:}  Suppose some $P$ satisfies the above condition. Then the symmetry of the identity channel under simultaneous permutation of the input and output alphabets means that we can always construct a new $P'$ which is symmetrised by the following `twirling' procedure
  \begin{equation*}
    P'(x,\hat{z}|z,y) = \frac{1}{|S_{g}|}\sum_{\pi \in S_g} P(x,\pi(\hat{z})|\pi(z),y)
  \end{equation*}
  where $S_g$ is the symmetric group of order $g$ and $\pi(z)$ is the image of $z$ under the permutation $\pi$. This clearly simulates the same channel as $P$, but it is highly symmetric in that
  \begin{equation*}
    P'(x,\hat{z}|z,y) =
    \begin{cases}
      D_{xy} & \text{ if } \hat{z} = z, \\
      Q_{xy} & \text{ if } \hat{z} \neq z.
    \end{cases}
  \end{equation*}

\dl{With this simplification in mind, we maximize $g$ such that a valid non-signalling correlation $P'$ allows the simulation of a $g$-message identity channel.  We enumerate the constraints on $P'$ in terms of $D$ and $Q$. } 
  
\noindent (1) $P'$ is a valid conditional probability distribution iff
  \begin{equation*}
    \forall x,y :  D_{xy} \geq 0,\ Q_{xy} \geq 0
  \end{equation*}
  and
  \begin{equation*}
    \forall y: \sum_{x\in X} (D_{xy} + (g-1)Q_{xy}) = 1.
  \end{equation*}

\noindent \dl{(2) The non-signalling condition from Bob to Alice is given by: 
  \begin{equation*} 
    \forall y : D_{xy} + (g-1)Q_{xy} = u_x 
  \end{equation*}
for some $u_x$, whereas} the condition that Alice cannot signal to Bob is
  \begin{equation*}
    \forall y : \sum_{x\in X} D_{xy} = \sum_{x\in X} Q_{xy} \,. 
  \end{equation*}
\noindent (3) The resulting channel is the $g$-message identity iff
  \begin{align*}
    & \sum_{x \in X, y \in Y} D_{xy} \mc{N}(y|x) = 1 \quad \text{and} \\
    & \sum_{x \in X, y \in Y} Q_{xy} \mc{N}(y|x) = 0.
  \end{align*}
\dl{Eliminating $D$ using condition (2), the full set of constraints (in terms of $Q$ and $u$) can be simplified:
  \begin{align*}
     \forall x,y: Q_{xy} \geq 0 \,, ~~ u_x \geq (g-1)Q_{xy},\\
	\forall y: \sum_{x\in X} Q_{xy} = \frac{1}{g}, \sum_{x\in X} u_x = 1\\
	~\mt{and}~ \sum_{x \in X, y \in Y} Q_{xy} \mc{N}(y|x) = 0 \,.
  \end{align*}
}


$\ozec{\NS}(\mc{N})$ is the largest integer smaller than the largest real number $g$ satisfying these constraints, which we now show is the $\alpha^{\ast}(H(\mc{N}))$ of the theorem. Defining $T_{xy} := (g-1)Q_{xy}$, the largest feasible value of $g$ is

  \begin{equation*}
    \begin{split}
      g = \max \bigg\{ &\frac{1}{1-s} : \sum_{x \in X} T_{xy} = s, T_{xy} \geq 0, u_y \geq T_{xy},\\
      &\sum_{x\in X} u_x = 1, \sum_{x \in X, y \in Y} T_{xy} \mc{N}(y|x) = 0 \bigg\}.
    \end{split}
  \end{equation*}
  By a simple application of the linear-fractional programming technique~\cite{boyd-book} this optimisation can be recast as a linear program: Making the substitutions $t := 1/(1-s)$, $T^\prime_{xy} := tT_{xy}$, $v(x) = t u_x$ yields
  \begin{equation*}
    \begin{split}
      g = \max \bigg\{ &t : v(x) \geq T^\prime_{xy} \geq 0, \\
          &\sum_{x\in X} v(x) = t,\;\; \sum_{x\in X} T^\prime_{xy} = t-1, \\
          &\sum_{\mathclap{x \in X, y \in Y}} T^\prime_{xy} \mc{N}(y|x) = 0
        \bigg\}.
      \end{split}
  \end{equation*}
  This is equivalent to the linear program
  \begin{equation*}
    \begin{split}
      g =\max\bigg\{& \sum_{x\in X} v(x) : v(x) \geq T^\prime_{xy} \geq 0,\\
         &\sum_{x \in X} (v(x) - T^\prime_{xy}) \leq 1,\\
		&\sum_{\mathclap{x \in X, y \in Y}} T^\prime_{xy} \mc{N}(y|x) = 0
       \bigg\}.
     \end{split}
  \end{equation*}
	The would just be a reorganisation, except that we have also replaced the equality constraints on the second line with inequalities. This doesn't change the value of the linear program: In any optimal solution with the inequalities, the sum over $v(x)$ will be at least one. Therefore, lowering the values of the $T^\prime_{xy}$, a solution to the LP where the equalities hold can be found which has the same objective value.
 
Finally, note that the $T^\prime_{xy}$ are redundant in the above formulation. Indeed, we may always set $T^\prime_{xy} = v(x)$ unless we are forced to take $T^\prime_{xy}=0$ due to $\mc{N}(y|x)>0$ or, equivalently, due to $\lceil\mc{N}(y|x)\rceil = 1$. Therefore,
  \begin{equation*}
    \begin{split}
      g =\max\bigg\{
        &\sum_{x\in X} v(x) : v(x) \geq 0 \ \forall x \in X,\\
        &\sum_{x} \lceil\mc{N}(y|x)\rceil v(x) \leq 1\ \forall y \in Y
      \bigg\},
    \end{split}
  \end{equation*}
  precisely the fractional packing number of $H(\mc{N})$.
\end{proof}


\begin{corollary}
  \label{cor:CNS-lower bound}
  If a channel $\mc{N}$ with $n$ inputs has at most $m$ non-zero entries $\mc{N}(y|x)$ for each $y$ (i.e.~the hyperedges of the equivocation graph are all of size $\leq m$). Then,
  \begin{equation*}
    \ozec{\NS}(\mc{N})
      \geq \left\lfloor \frac{n}{m} \right\rfloor
    \text{ and }
    \zec{\NS}(\mc{N})
      \geq \log \frac{n}{m}.
  \end{equation*}
  The \emph{proof} is by checking that the assignment $v(x) = \frac{1}{m}$ is feasible. \qed
\end{corollary}

We now show that $\zec{{\NS}}$ can be arbitrarily larger than $\zec{~}$. In fact, there are channels for which the latter is $0$ while the former is positive! 

\dl{Let $\binom{[n]}{m}$ denote the set of all size-$m$ subsets of $[n]$.} 
For all $n > m \geq 2$, define the channels $\mc{S}_{n,m} \in \CCs{[n]}{\binom{[n]}{m}}$, each maps $x \in [n]$ to a random subset of $[n]$ of cardinality $m$ containing $x$, i.e.
\begin{equation*}
  \mc{S}_{n,m}(y|x) =
  \begin{cases}
    0 & \text{ if } x\not\in y, \\
    \frac{1}{{n-1 \choose m-1}} & \text{ if } x \in y.
  \end{cases}
\end{equation*}

For all these channels, $\ozec{}(\mc{S}_{n,m}) = \zec{}(\mc{S}_{n,m}) = 0$ because any two inputs $x$ and $x'$ are contained in a common set, hence they are confusable. (In other words, the confusability graph is the complete graph $K_n$.)
On the other hand, by Corollary \ref{cor:CNS-lower bound}, all of these channels have $\zec{\NS}(\mc{S}_{n,m}) \geq \log \frac{n}{m}$, \dl{a strictly positive non-signalling-assisted capacity}. 

The smallest parameters for which this effect can be seen on the single-shot level are $n=4$ and $m=2$: $\mc{S}_{4,2}$ is a channel with $4$ inputs and $6$ outputs, and Theorem~\ref{NSZELP} gives $\ozec{\NS}(\mc{S}_{4,2}) = 1$. How can this be? Define a non-signalling correlation $P \in \aBCC{\NS}{\{0,1\}}{[4]}{\binom{[4]}{2}}{\{0,1\}}$ as follows. Alice's input is a bit $z$, her output $x'$ is a random element of $[4]$. Bob's input is a subset $y' \in \binom{[4]}{2}$. If $x' \in y'$ then Bob's output bit $\hat{z}$ is $z$ and otherwise is $\lnot z$. Clearly, Bob's output is independent of Alice's input and vice versa so it is indeed non-signalling.

Suppose Alice wires her output into the channel $\mc{S}_{4,2}$ (so $x' = x$) and Bob uses the output of $\mc{S}_{4,2}$ as his input to $P$ (so $y' = y$). The behaviour of the channel ensures that $y'$ will always contain $x'$ and therefore Bob's output $\hat{z}$ will always be equal to $z$. A bit is transmitted from Alice to Bob with perfect reliability --- and that despite the fact that any two inputs of the channel cannot be told apart with certainty by Bob!

Whenever $\zec{}(\mc{N}) > 0$, the non-signalling assisted zero-error capacity $\zec{\NS}(\mc{N})$ is precisely the same as the feedback assisted zero-error capacity. This is especially remarkable because the corresponding quantities for a finite number of channel uses are not necessarily the same, and the proofs of the capacity formulas are very different \cite{shannon}. Also interesting is the fact that non-signalling proves strictly more powerful than feedback. In fact, when the capacities differ, the feedback assisted capacity must be zero.


\subsection{Assistance by entanglement}
In \cite{CLMW} we show that, like $\ozec{}(\mc{N})$ and unlike $\ozec{\NS}(\mc{N})$, the one-shot (and hence also, asymptotic) entanglement assisted zero-error capacity depends only on $G(\mc{N})$. An immediate corollary of this is that if $\ozec{}(\mc{N}) = 0$ then $\ozec{\SE} = 0$. Proposition~\ref{pw-G} of the appendix, shows that these facts hold for assistance by any class of correlations with a certain operational property, which is possessed by $\SE$ but not $\NS$. Here, for the reader's convenience we repeat the proof of \cite{CLMW}.


\begin{theorem}\label{SEA}
	For any channel $\mathcal{N}$ with inputs $X$ and outputs $Y$, \dl{$\ozec{\SE}({\mathcal N}) = \max c$ subject to the constraint} that there exists a density matrix $\rho_B$ and positive semidefinite operators $\beta^{(z)}_x$ for all $z \in [c]$, $x\in X$, on some Hilbert space such that,
	\begin{gather*}
		\forall z : \sum_{x\in X} \beta^{(z)}_x = \rho_B\\
		\forall z\neq z', \{x,x'\} \in E(G(\mc{N})) : \tr \beta^{(z)}_x \beta^{(z')}_{x'} = 0.
	\end{gather*}
	Consequently, $c_{\SE}(\mathcal{N})$ depends only on $G(\mc{N})$.
\end{theorem}
\begin{proof}
We call the shared entangled state $\rho_{AB}$. Without loss of generality, to send message $z$, Alice performs a measurement with POVM elements $\{M^{(z)}_x : x\in X\}$, and with probability $p^{(z)}_x (= \!\! \tr [ M^{(z)}_x ( \tr_B \rho_{AB} ) ] \;)$, obtains outcome $x$. Conditional on the knowledge of $z$ and $x$, the residual state of Bob's system is $\rho^{(z)}_x = (\tr_A M^{(z)}_x \ox \openone \rho_{AB})/p^{(z)}_x$. Letting $\beta^{(z)}_x := p^{(z)}_x \rho^{(z)}_x$, for all messages $z$
\[
	\sum_x \beta^{(z)}_x = \tr_A \rho_{AB} =: \rho_B
\]
reflecting the fact that without information from the classical channel, Bob has no idea which message Alice sent (i.e.\ causality is respected). Conversely, any set of positive operators $\beta^{(z)}_x$ which satisfy this condition for some $\rho_B$ can be realised by a suitable choice of $\rho_{AB}$ and generalised measurements.

Alice puts the outcome $x$ into the channel $\mc{N}$. \dl{Bob obtains
the channel output $y$, in addition to a quantum state left in his
half of the entangled system.  This bipartite state on Bob's side is given by:}
   \begin{equation*}
     \sigma_z := \sum_{x\in X,y\in Y} \mc{N}(y|x) \proj{y} \ox \beta^{(z)}_x.
   \end{equation*}
   The encoding works if and only if Bob can distinguish perfectly between all the $\sigma_z$, i.e.\ for all distinct $z, z' \in [c]$
\begin{equation*}
     \begin{split}
       0	&= \tr \sigma_z \sigma_{z'}\\
		&= \sum_{x,x' \in X, y,y'\in Y} \mc{N}(y|x)\mc{N}(y'|x') \delta_{yy'} \tr \beta^{(z)}_x \beta^{(z')}_{x'} \\
       	&=\sum_{x,x':\{x,x'\}\in E(G)} \sum_y \mc{N}(y|x)\mc{N}(y|x') \tr \beta^{(z)}_x \beta^{(z')}_{x'}.
     \end{split}
   \end{equation*}
  \end{proof}


Entanglement can still help, though. As shown in Theorem~\ref{KSchan} (and previously in \cite{CLMW}) there are channels with $\ozec{\SE}(\mc{N}) > \ozec{}(\mc{N}) > 0$.

Whether there are channels $\mc{N}$ exhibiting an asymptotic separation $\zec{\SE}(\mc{N}) > \zec{}(\mc{N})$ remains an open question at this time. The efficiently computable formulae for $\ozec{\NS}$ and $\zec{\NS}$ derived in the previous section provide upper bounds on entanglement assistance in both the one shot and asymptotic cases, but a tighter bound is known: Duan \emph{et al.} \cite{DSW} have defined a generalisation of the Lov\'{a}sz theta function \cite{Lovasz} for quantum channels. It is multiplicative for tensor products of channels and reduces to the classical Lov\'{a}sz theta function when the channel is classical. They show that this function is an upper bound on the entanglement assisted one-shot zero-error capacity (for sending classical messages) for any \emph{quantum} channel. Therefore, the classical theta function is an upper bound on $\ozec{\SE}$ for classical channels. A short and direct proof of this fact was derived independently by Beigi \cite{beigi}. The bound is a strict improvement over the fractional packing bound and since $\vartheta$, like $\alpha^\ast$, is multiplicative, it too can be immediately applied to the asymptotic rate:
\[
 	\zec{\SE}(\mc{N}) \leq \log \vartheta(G(\mc{N})).
\]

In terms of trying to decide whether separations exist between $\zec{\SE}$ and $\zec{}$, this is a rather frustrating result because $\vartheta(G(\mc{N})$ is typically also the best bound we have on $\zec{}$! Exceptions to this have been found, by Haemers \cite{Haemers}, but then the problem is to determine whether entanglement assisted protocols exist which beat the best known upper bound on $\zec{}$ for those special cases, and even then, only a positive answer would settle the general problem.

Another intriguing corollary of the result is that for the channels with $\ozec{} < \ozec{\SE}$ made according to our construction from \cite{CLMW}, the Lov\'{a}sz theta function coincides exactly with the \emph{lower} bound on $\ozec{\SE}$ provided by the explicit protocol we give, so for these channels we know the precise value of $\zec{\SE}(\mc{N})$ and furthermore that it is achieved by repeating the optimal protocol for a single use of the channel.

We now review the proof of the above statement as well as the construction from \cite{CLMW} to which it applies.


\begin{definition}
	Let $G$ be a graph with vertex set $X$. An \emph{orthonormal representation} $\Gamma$ of $G$ in $\mathbb{C}^d$ is an assignment of unit vectors in $\mathbb{C}^d$ to the vertices of $G$ such that \dl{if two vertices connected by an edge then their assigned vectors are orthogonal} (where orthogonality is with respect to the usual inner product $\langle \cdot, \cdot \rangle$):
	\[
		\forall x,x' \in X: \langle\Gamma(x),\Gamma(x')\rangle \iff \{x,x'\} \in E(G).
	\]
\end{definition}

\begin{theorem}\label{partG}
	Suppose that $G$ is a graph with an orthonormal representation in $\mathbb{C}^d$ whose vertices can be partitioned into exactly $q$ cliques $\{\mathcal{K}_1, \ldots \mathcal{K}_q\}$ each of size $d$.  \dl{Then there is a one-shot zero-error communication protocol assisted by a rank-$d$ maximally entangled state, which shows that $\ozec{\SE}(G) \geq q$.  Also, $\vartheta(G) = q$, and since \cite{beigi,DSW} proved that $\ozec{\SE}(G) \leq \vartheta(G)$.  Therefore, $\ozec{\SE}(G) = q$.} 

\end{theorem}
%

\begin{proof}
First, we describe the entanglement assisted protocol. Alice and Bob
share $\frac{1}{\sqrt{d}}\sum_{j = 1}^{d} \ket{j}_A\ox\ket{j}_B$, with
$\ket{j}$ the computational basis vectors for each party.  The $q$
cliques of size $d$ which partition the vertices of the graph
correspond to $q$ complete orthonormal bases for $\mathbb{C}^d$
\dl{given by ${\mathcal B}_z = \{ (\Gamma{(x)}: \forall x \in
{\mathcal K}_z \}$ for $z = 1$ to $q$.  To encode the message $z$,
Alice measures her half of the shared state along the basis ${\mathcal
B}^c_z$ (obtained by conjugating each state in ${\mathcal B}_z$).  If
the outcome corresponds to $x$, Bob's subsystem is left in the state
$\Gamma(x)$.}  \footnote{\dl{If $\Gamma{(x)} = \sum_i a_i \ket{i}$, then, 
the postmeasurement state is $(\sum_i a_i \bra{i}_A \otimes I_B) \sum_{j =
1}^{d} \ket{j}_A\ox\ket{j}_B = \sum_j a_j \ket{j}_B = \Gamma{(x)}_B$.}}

	Alice inputs $x$ to the channel. Bob's output $y$ from $\mc{N}$ tells him \dl{a clique $e_y$ in $G$ that contains $x$, which is not necessarily one of cliques in the partition.  So Bob's subsystem must be} in one of the corresponding set of \emph{mutually orthogonal} states $\Gamma(e_y)$. Therefore, he can perform a projective measurement on his subsystem to determine exactly which state he has, from which he can deduce $x$ and, \emph{a fortiori}, the symbol $z \in [q]$ which Alice chose, with certainty.

Second, to obtain $\vartheta(G)$, note that it can only increase if edges are removed, and it is multiplicative under strong graph product we have
	\[
		\vartheta(G) \leq \vartheta(\bar{K}_q\ox K_d) = \vartheta(\bar{K}_q)\vartheta(K_d) = q,
	\]
	where $K_n$ and $\bar{K}_n$ are the complete and empty graphs on $n$, which have Lov\'{a}sz theta values of $1$ and $n$, respectively.
	
%

\dl{Using the result from \cite{beigi,DSW} that $\ozec{\SE}(G) \leq \vartheta(G)$, and putting both parts together,} 
\[
	\ozec{\SE}(G) = \vartheta(G) = q.
\]
\end{proof}

\begin{definition}\label{KSset}
We call a set $Z = \{B_m\}_{m=1}^q$ of $q$ complete orthogonal bases $B_m = \{\ket{b_{mj}}:j=1,\ldots,d\}$ for $\mathbb{C}^d$ a \emph{KS basis set}, if it is impossible to pick one vector from each basis so that no two are orthogonal.
\end{definition}

That such sets exist is a simple corollary of the Kochen-Specker theorem \cite{KS}. An example of a KS basis set with 6 bases for $\mathbb{C}^4$ taken from a proof of the Kochen-Specker theorem by Peres \cite{Peres-KS} is given in \cite{CLMW}.

\begin{theorem}\label{KSchan}
	For any KS basis set $Z = \{B_m\}_{m=1}^q$ in $\mathbb{C}^d$ of $q$ bases, one can construct a classical channel $\mathcal{N}_Z$ (with $q d$ input symbols) with $c_{0}(\mathcal{N}_Z) < q$ and $c_{\SE}(\mathcal{N}_Z) = q$.
\end{theorem}
\begin{proof}
	Construct the graph $G_{Z}$ on $[q]\times[d]$ with $(m,j)$ connected to $(m',j')$ iff $\ket{b_{mj}}$ and $\ket{b_{m'j'}}$ are orthogonal. Clearly, $G_{Z}$ partitions into $q$ cliques corresponding to the $q$ bases in $Z$, so $\alpha(G) \leq q$, and if there was an independent set in $G_{z}$ of size $q$, it would have to have one element in each of the $q$ cliques. But this would correspond to a selection of one vector in each basis in $Z$ such that no two are orthogonal, in contradiction to the fact that $Z$ is a KS basis set.

	Letting $\mc{N}_Z$ be a channel with confusability graph $G_Z$, we have just shown that $\ozec{}(\mc{N}_Z) < q$. On the other hand, since $\Gamma((m,j)) := \ket{b_{mj}}$ clearly defines an orthonormal representation of $G_{Z}$ in $\mathbb{C}^d$, Theorem~\ref{partG} tells us that $\ozec{\SE}(\mc{N}_{Z}) = q$ (and that this can be achieved using a rank-$d$ maximally entangled state).
\end{proof}

\section{Exactly simulating noisy channels with perfect communication}
\label{sec:sim}
This section concerns the ``reverse'' problem to the zero-error channel coding problem: How much zero-error communication is \emph{required} to exactly simulate a \emph{noisy} channel. It will turn out that the one-shot communication cost can differ wildly between availability of no correlation, shared randomness, entanglement and non-signalling resources. However, in the many-copy limit they all turn out to give the same rate, as long as shared randomness is available. Under a relaxed (namely: combinatorial) notion of channel simulation, we find complete reversibility between non-signalling assisted zero-error channel coding and channel simulation.

\subsection{Without any assistance}
What does it mean to simulate a channel $\mc{N} \in \CCs{X}{Y}$? With a $k$ symbol identity channel and no other correlations between sender and receiver the most general protocol is simply this: Alice applies a local channel $\mc{Q} \in \CCs{X}{[k]}$ and sends to Bob the result, on which he applies a channel $\mc{R} \in \CCs{[k]}{Y}$. The composition should be the desired channel $\mc{N} = \mc{R} \circ \mc{Q}$.

\begin{theorem}
  \label{thm:no-resources-sim}
  For a channel $\mc{N} \in \CCs{X}{Y}$, $\oxk{}(\mc{N})$ equals the \emph{positive-rank} \cite{positive-rank} of the transition probability matrix $\mc{N}(y|x)$, i.e. the smallest number $k$ of probability distributions on $Y$ such that their convex hull contains all of the output distributions $\mc{N}(\cdot|x)$.

  Since positive-rank is lower bounded by linear rank, we get the following lower bounds:
  \begin{equation*}
    \oxk{}(\mc{N}) \geq \rank\mc{N}, \quad K_0(\mc{N}) \geq \log\rank\mc{N},
  \end{equation*}
  the latter because the rank is multiplicative.  \qed
\end{theorem}

For instance, the channel $\mc{N}_{\rm{NOT}}\in\CCs{[n]}{[n]}$ with $\mc{N}_{\rm{NOT}}(y|x) = 0$ if $y=x$ and $\frac{1}{n-1}$ if $y \neq x$, will have $K_0(\mc{N}_{\rm{NOT}}) = \log n$, the same as the perfect channel, even though both its Shannon and zero-error capacities are much lower.

\subsection{With shared randomness}

The set of channels one can perfectly simulate by sending one of $k$ symbols when arbitrary shared randomness is available is simply the convex hull of the set (just described) that can be achieved without shared randomness:
\begin{theorem}\label{kSR}
  For a channel $\mc{N} \in \CCs{X}{Y}$, $\oxk{\SR}(\mc{N})$ is the minimum integer $k$ such that
  \begin{equation*}
    \mc{N} \in \operatorname{conv} \left(
      \bigcup_{Z \subseteq Y,\ |Z| \leq k} \CCs{X}{Z} \right),
  \end{equation*}
  where we view $\CCs{X}{Z}$ naturally as a subset of $\CCs{X}{Y}$.
	In fact, on the right hand side, we may replace the sets $\CCs{X}{Z}$ with their corresponding subsets of deterministic channels. As matrices, these are zero/one stochastic matrices, with rank $\leq k$ and a channel $\mc{N}$ has $\oxk{\SR}(\mc{N}) \leq k$ iff its matrix is a convex combination of these.
\end{theorem}
\begin{proof}
  Any protocol to simulate $\mc{N}$ exactly using shared randomness and $k$ messages amounts to writing $\mc{N}$ as a convex (probability) combination of product channels,
  \begin{equation*}
    \mc{N} = \sum_i p_i \mc{R}_i \circ \mc{Q}_i,
  \end{equation*}
  with $\mc{Q}_i \in \CCs{X}{[k]}$, $\mc{R}_i \in \CCs{[k]}{Y}$, and $p_i \geq 0, \sum_i p_i = 1$. Since the extreme points in the set of channels are the deterministic channels, we may push the randomness involved in forming a stochastic map into the shared randomness. But since $\mc{Q}_i$ has only $k$ outputs symbols and $\mc{R}_i$ is deterministic, also the composition $\mc{R}_i \circ \mc{Q}_i$ can have only $k$ possible output symbols forming a subset $Z\subseteq Y$, $|Z| \leq k$. In other words, $\mc{N}$ is a convex combination of deterministic channels in $\CCs{A}{Z}$, for $k$-element subsets $Z \subseteq Y$.
\end{proof}

\subsection{Non-signalling correlations}
Just as in the case of zero-error communication, making non-signalling correlations freely available gives a very tractable structure to the problem of perfectly simulating noisy channels and the one-shot communication cost $\oxk{\NS}(\mc{N})$ has a correspondingly simple form: It is the smallest integer greater than or equal to a certain simple norm of the conditional probability matrix. This norm is multiplicative under tensor products, so the corresponding asymptotic rate $\xk{\NS}(\mc{N})$ is just its logarithm.

\begin{theorem}\label{NSrate}
  For a channel $\mc{N} \in \CCs{X}{Y}$,
  \begin{equation*}
    \oxk{\NS}(\mc{N})
    = \left\lceil\sum_y \max_x \mc{N}(y|x)\right\rceil,
  \end{equation*}
  and since $\sum_y \max_x \mc{N}^{\otimes n}(y|x) = (\sum_y \max_x \mc{N}(y|x))^n$, (note that this function is a norm on stochastic matrices), the corresponding asymptotic rate is just
  \begin{equation*}
    \xk{\NS}(\mc{N}) 
    =  \log \left( \sum_y \max_x \mc{N}(y|x) \right).
  \end{equation*}
\end{theorem}
\begin{proof}
  If $\mc{N}$ is in $\CCs{X}{Y}$, it can be simulated with a $k$-input identity channel and non-signalling correlations if and only if there exists $P$ in $\aBCC{\NS}{X}{[k]}{[k]}{Y}$ such that
  \begin{equation}\label{contracts2chan}
    \sum_{z = \hat{z}} P(z,y|x,\hat{z}) = \mc{N}(y|x).
  \end{equation}
  Again the twirling procedure in the proof of Theorem~\ref{NSZELP} simplifies things, but now the symmetry is in the identity channel used for the simulation. Defining
  \begin{equation*}
    P'(z,y|x,\hat{z}) = \frac{1}{|S_{k}|}\sum_{\pi \in S_k} P(\pi(z),y|x,\pi(\hat{z}))
  \end{equation*}
  where $S_k$ is the symmetric group of order $k$ and $\pi(z)$ is the image of $z$ under the permutation $\pi$ yields a correlation which simulates the same channel when used in place of $P$ (summing over $\pi(z) = \pi(\hat{z})$ is the same as summing over $\hat{z} = z$), but where
\begin{equation*}
    P'(z,y|x,\hat{z}) =
    \begin{cases}
      D_{yx} & \text{ if } \hat{z} = z, \\
      Q_{yx} & \text{ if } \hat{z} \neq z.
    \end{cases}
  \end{equation*}

\dl{We now list the conditions on $P'$ (in terms of $D$ and $Q$): }

\noindent (1) The correctness of the simulation is given by 
Eq.~\eqref{contracts2chan}: 
  \begin{equation*}
    D_{yx} = \mc{N}(y|x)/k.
  \end{equation*}

\noindent (2a) The conditions for no signalling from Alice to Bob are
  \begin{equation*}
    \sum_{z\in [k]} P'(z,y|x,\hat{z}) = \sum_{z\in [k]} P'(z,y|x',\hat{z})
  \end{equation*}
  for all $x,x'$, which reduce to
  \begin{equation*}
    D_{yx} + (k-1)Q_{yx} = D_{yx'} + (k-1)Q_{yx'},
  \end{equation*}
  so we write $D_{yx} + (k-1)Q_{yx} = u_y$. Clearly we require that $\sum_y u_y = 1$ and $u_y \geq 0$ (in fact, $u_y$ is just the marginal distribution of the output $y$ which is independent of both inputs, like in the PR box).

\noindent (2b) The conditions for no signalling from Bob to Alice,
  \begin{equation*}
    \dl{ 
    \forall \hat{z},\hat{z}',x : \sum_y P'(z,y|x,\hat{z}) 
       = \sum_y P'(z,y|x,\hat{z}') \,,
    }
  \end{equation*}
  which reduce to $\sum_y D_{yx} = \sum_y Q_{yx} \forall x$,
	are already ensured by the condition that $D_{yx} =
  \mc{N}(y|x)/k$ and $\sum_y D_{yx} + (k-1)Q_{yx} = 1$ for all $x$
  (these mean that $\sum_y D_{yx} = \sum_y Q_{yx} = 1/k$ for all $x$).

\noindent (3) The only other constraint is that the entries of $Q$ are positive. 

Putting these constraints together, we see that a suitable $P'$ (and hence $P$) exists if and only if there is a probability vector $u$ such that the resulting $Q$ matrix has positive entries, i.e.
  \begin{equation*}
    u_y - \mc{N}(y|x)/k \geq 0
  \end{equation*}
  for all $y,x$. Such a $u$ is possible if and only if $\displaystyle\sum_y \max_x \mc{N}(y|x)/k \leq 1$.
\end{proof}


\begin{remark}
  It is not hard to verify directly that the bit rate needed to perfectly simulate $\mc{N}$ with free non-signalling correlations is greater than the Shannon capacity of $\mc{N}$. If the channel input is the random variable $R_x$ where $\Pr (R_x = x) = p(x)$ and the resulting channel output is the random variable $R_y$ then
  \begin{align*}
    I(R_x:R_y) &= \sum_{x,y} \mc{N}(y|x)p(x) \log\frac{\mc{N}(y|x)}{\sum_z \mc{N}(y|z) p(z)} \\
    &\leq \log\sum_{x,y}
      \frac{\mc{N}(y|x)^2 p(x)}{\sum_z \mc{N}(y|z) p(z)}\\
    &\leq \log\sum_{x,y}
       \frac{\mc{N}(y|x) p(x) \max_{r} \mc{N}(y|r)} {\sum_z \mc{N}(y|z) p(z)}\\
    &=\log\sum_{y} \max_{r} \mc{N}(y|r).
  \end{align*}
  The Shannon capacity of $\mc{N}$ is obtained by maximising the left-hand side over all input distributions $p$.
\end{remark}

\subsection{Arbitrarily large gap between $\oxk{}$, $\oxk{\SR}(\mc{N})$, and 
$\oxk{\NS}(\mc{N})$}

Shared randomness is one type of non-signalling correlation so it is clear that $\oxk{\SR}(\mc{N}) \geq \oxk{\NS}(\mc{N})$.  \dl{It turns out that there can be an arbitrarily large gap between these two costs.  This is the case for the ``universal channels'' to be defined below.} 

\begin{definition}\label{Uchans}
  \dl{Recall that the set of all size-$m$ subsets of $[n]$ is denoted 
  by ${[n] \choose m}$.}  The \emph{universal channel} $\mc{U}_{n,m}$ is the channel in $\CCs{{[n] \choose m}}{[n]}$ with,
  \begin{equation*}
    \mc{U}_{n,m}(y|x) =
    \begin{cases}
      \frac{1}{m} & \text{ if } y\in x, \\
      0           & \text{ if } y\not\in x.
    \end{cases}
  \end{equation*}
  In words, the channel takes as input a set $x \in {[n] \choose m}$ and outputs a random element of that set.
\end{definition}
\dl{Note that $\mc{N}_{\rm{NOT}}$ introduced earlier in this section is a
special case of the universal channel with $m=n-1$.}

\dl{The universal channels} have a great deal of symmetry. The symmetric group $S_n$ acts on both the input and the output alphabet of $\mc{U}_{n,m}$: on the latter naturally as permutations of the symbols (written $\pi(y)$), on the former as simultaneous permutations of all elements in the sets $x \subseteq [n]$ (written $x^\pi$). With these actions, $\mc{U}_{n,m}$ is $S_n$-covariant:
\begin{equation}
  \label{eq:covariance}
  \forall y,x \quad \mc{U}_{n,m}(y|x) = \mc{U}_{n,m}(\pi(y)|x^\pi).
\end{equation}
A beautiful consequence of the covariance is that it specifies $\mc{U}_{n,m}$ almost uniquely: $\mc{U}_{n,m}$ is the only channel satisfying eq.~\eqref{eq:covariance} and in addition $\mc{U}_{n,m}(y|x) = 0$ if $y\not\in x$.

To simulate $\mc{U}_{n,m}$ with zero error when assisted by arbitrary non-signalling correlations, Theorem~\ref{NSrate} shows one needs a noiseless channel of
\begin{equation*}
  \oxk{\NS}(\mc{U}_{n,m})
  = \left\lceil \sum_y \max_x N(y|x) \right\rceil
  = \left\lceil \frac{n}{m} \right\rceil
\end{equation*}
many symbols, and this is sufficient. The minimal asymptotic rate of communication needed given free non-signalling correlations is $\log\frac{n}{m}$. On the other hand, when only shared randomness is available, the communication cost can be much higher:

\begin{proposition}
\label{ksr}
  For any $n \geq m \geq 1$,
  \begin{equation*}
    \oxk{\SR}(\mc{U}_{n,m}) = n-m+1.
  \end{equation*}
\end{proposition}
\begin{proof}
  \dl{We first show that $k=n-m$ is not sufficient by contradiction.  Recall Theorem~\ref{kSR} and} consider an element $\mc{N}$ from $\CCs{X}{Z}$, with $Z \subseteq Y$, $|Z|=n-m$ in the convex decomposition of $\mc{U}_{n,m}$ that comes with a strictly positive weight $p$. That means, for any input $x$,
  \begin{equation*}
    p \mc{N}(\cdot|x) \leq \mc{U}_{n,m}(\cdot|x)
  \end{equation*}
  in the sense of element-wise ordering of the probability vectors. Choosing $x = [n]\setminus Z$ --- which has cardinality $m$ --- leads to the desired contradiction: restricted to $Z$, $\mc{U}_{n,m}(\cdot|x)$ is the zero vector (see the definition), whereas $\mc{N}(\cdot|x)$ has all of its probability mass in $Z$.

 \dl{On the other hand, there is a protocol that uses only $n-m+1$ messages:} The shared randomness is a uniformly distributed subset $T \in {[n] \choose n-m+1}$. On input $x$, Alice selects a uniformly random element $y \in x \cap T$, which is non-empty by the pigeonhole principle. To send $y$ to Bob, she needs only a number from $1$ through $n-m+1$ to specify where $y$ occurs in $T$. Clearly, this protocol, and hence the simulated channel, has the same $S_n$-covariance as $\mc{U}_{n,m}$. Furthermore, the simulated channel assigns zero conditional probability to all $y \not\in a$ for input $x$. Thus, the simulated channel must be $\mc{U}_{n,m}$.
\end{proof}

\dl{These universal channels provide simple and highly structured examples for separating $\oxk{}$, $\oxk{\SR}$, and $\oxk{\SR}$.  We have already seen that for $m=n-1$, $\oxk{}(\mc{U}_{n,m}) = n$ but Prop.~\ref{ksr} says that $\oxk{\SR}(\mc{U}_{n,m}) = 2$.  In a different regime, for example when $n$ is even and $m=\frac{n}{2}$, $\oxk{\SR}(\mc{U}_{n,m}) = \left(\frac{n}{2}+1\right)$ but $\oxk{\NS}(\mc{U}_{n,m}) = 2$.  In both cases, the separation is of order $n$ which is maximal given the size of the input alphabet.}   



\subsection{Shared entanglement}
The possibility of large separations between $\oxk{\NS}(\mc{N})$ and $\oxk{\SR}(\mc{N})$, raises the question of where the power of the intermediate shared entanglement class fits in between the two. While we do not have a general understanding of this matter yet, we can at least give examples where entanglement beats shared randomness and cases where general non-signalling correlations can beat entanglement.

Let $\mc{T}_p$ denote the ternary erasure channel with transmission probability $p$:
\begin{equation*}
  \mc{T}_p = \left(
    \begin{array}{ccc}
      p & 0 & 0 \\
      0 & p & 0 \\
      0 & 0 & p \\
      1-p & 1-p & 1-p
    \end{array}
  \right).
\end{equation*}

We can use the ideas of the appendix to show that non-signalling correlations can beat shared entanglement for one-shot simulation of $\mc{T}_{1/2}$.
\begin{proposition}
  Whereas $\oxk{\NS}(\mc{T}_{1/2}) = 2$, $\oxk{\SE}(\mc{T}_{1/2}) = 3$. Therefore, using a single (perfect) bit of communication, strictly more channels can be simulated if generalised non-signalling correlations are available rather than entanglement.
\end{proposition}
\begin{proof}
	By using the twirling procedure of Theorem~\ref{NSrate} (which can be assumed w.l.o.g. whenever shared randomness is available), and considering the simplified non-signalling constraints which result, it is not hard to see that exact simulation of $\mc{T}_{1/2}$ with a single bit is equivalent to the ability to realise a particular non-signalling correlation $P^{\star}$ defined, in the notation of Theorem~\ref{NSrate}, by $k = 2$,
	\begin{equation*}
	  D_{yx} = \frac{1}{4}\left(
	    \begin{array}{ccc}
	      1 & 0 & 0 \\
	      0 & 1 & 0 \\
	      0 & 0 & 1 \\
	      1 & 1 & 1
	    \end{array}
	  \right)
	\end{equation*}
	and $Q_{yx} = 1/4 - D_{yx}$. This $P^{\star}$ must be in the class of available correlations. Applying the observations of the appendix, and looking at the 8 conditional channels $P^{\star}_{xz}$, one finds that they are all pair-wise distinguishable and therefore $P^{\star}$ cannot be in the class $\SE$, by Proposition~\ref{pw-bound}.
\end{proof}

\begin{proposition}
  \dl{Strictly more classical channels can be simulated using shared entanglement than can with shared randomness.  In particular, there are channels $\mc{N}$ for which $\oxk{\SE}(\mc{N}) = 2$ but $\oxk{\SR}(\mc{N}) \geq 3$.} 
\end{proposition}
\begin{proof}
We construct these channels and demonstrate the separation in three steps.  First, we show that $\mc{T}_{1/2}$ can be simulated using one bit of communication and a certain non-signalling correlation called the ``PR-box''.  Then, using the same protocol, but replacing the PR-box by a weaker correlation obtainable via shared entanglement, we write down the channel $\mc{N}$ that is being simulated.  Finally, we explain why $\mc{N}$ cannot be simulated with shared randomness and one bit of communication.  


The PR-Box (introduced by Popescu and Rohrlich \cite{PR}) is a particular correlation $P(s,t|a,b)$ given by: 
\begin{equation*}
    \bordermatrix{\text{}&(0,0)&(0,1)&(1,0)&(1,1)\cr
      (0,0)& 1/2   & 1/2 		& 1/2 		& 0 \cr
      (0,1)& 0 	& 0 	& 0 	& 1/2 \cr
      (1,0)& 0 	& 0 	& 0 	& 1/2 \cr
      (1,1)& 1/2 		& 1/2 		&1/2 		&0}
\end{equation*}
In other words, the outputs $s,t$ are random bits except for the constraint $s \oplus t = a \cdot b$.   We let $z$ to be the message sent via the classical channel.  

We first give an explicit method for simulating $\mc{T}_{1/2}$ using a bit of communication and a single use of a PR box.  Bob always chooses his PR-box input $b$ to be the channel output $z$ and outputs $(z,t)$ for the simulation.  
Alice has a message $x \in \{0,1,2\}$. 
If $x=0$, she sets her PR-box input to be $a=0$.  The PR-box outputs a random bit $s$.  
She sends $z=s$ to Bob, who puts it in the PR-box and obtains $t=s$.  
Thus, Bob outputs $(0,0)$ or $(1,1)$ randomly. 
If $x=1$, Alice sets $a=1$ and her PR-box outputs a random bit $s$.  
She chooses $z=s$.  For either value of $s$, $t=0$.  
Thus, Bob outputs $(0,0)$ or $(1,0)$ randomly. 
If $x=2$, Alice sets $a=0$ and $z=0$.  
Thus, Bob outputs $(0,0)$ or $(0,1)$ randomly.  
Identifying the outputs $(0,0)$, $(1,1)$, $(1,0)$ and $(0,1)$ with the erasure symbol {\sc e}, $0$, $1$, and $2$ respectively, the above simulates $\mc{T}_{1/2}$ perfectly.  

In the second step, we generalize the PR-Box to correlations $P_\lambda$ given by: 
\begin{equation*}
    \bordermatrix{\text{}&(0,0)&(0,1)&(1,0)&(1,1)\cr
      (0,0)& \lambda/2 		& \lambda/2 		&\lambda/2 		&\bar{\lambda}/2 \cr
      (0,1)& \bar{\lambda}/2 	& \bar{\lambda}/2 	&\bar{\lambda}/2 	&\lambda/2 \cr 
      (1,0)& \bar{\lambda}/2 	& \bar{\lambda}/2 	&\bar{\lambda}/2 	&\lambda/2 \cr
      (1,1)& \lambda/2 		& \lambda/2 		&\lambda/2 		&\bar{\lambda}/2},
  \end{equation*}
where $\bar{\lambda} = 1 - \lambda$.   Note that $P_1$ is the PR-Box.
The PR box can be approximated using a maximally entangled pair of qubits. An optimal approximation, in terms of the CHSH violation, yields the correlation $P_\lambda$ with $\lambda = (1+1/\sqrt{2})/2 \approx 0.85$. If this entanglement based approximation of the PR box is substituted into the protocol given above, the resulting channel ${\mathcal N}$ is given by 
  \begin{equation*}
    \bordermatrix{\text{}&0&1&2\cr
      \text{\sc e} \leftrightarrow (0,0)& \alpha & \alpha & 1/2 \cr
      0 \leftrightarrow  (1,1)& \alpha & \beta  & 0 \cr
      1 \leftrightarrow (1,0)& \beta & \alpha & 0 \cr
      2 \leftrightarrow (0,1)& \beta & \beta & 1/2},
  \end{equation*}
  where $\alpha = (1+1/\sqrt{2})/4 \approx 0.43$ and $\beta = 1/2 - \alpha \approx 0.07$.  

Finally, we check with a computer that ${\mathcal N}$ is not a convex combination of rank-two, zero-one, stochastic matrices and so, according to Theorem~\ref{kSR}, can't be exactly simulated using one bit of communication if only shared randomness is available.
\end{proof}

\subsection{Asymptotic equality of correlation assisted communication costs}
Among the results of this section so far are channels proving separations between the communication costs of NS-, SE- and SR-assisted channel simulation for a single channel use. In the case of NS vs. SR, the universal channels of Definition~\ref{Uchans} show that this gap can be arbitrarily large. Despite this, \dl{we will prove in Theorem \ref{srns}} that when simulating many uses of a channel, a protocol using shared randomness can achieve an \emph{asymptotic} rate of communication as low as the optimal rate with non-signalling assistance derived in Theorem~\ref{NSrate}. Since the rate with entanglement assistance is sandwiched between these two rates, it follows that $\xk{\SR}(\mc{N}) = \xk{\SE}(\mc{N}) = \xk{\NS}(\mc{N}) = \log \sum_y \max_x \mc{N}(y|x)$, for all channels $\mc{N}$.


\dl{The proof is structured into two steps.  First, we show that the
asymptotic equality discussed above holds for all the universal
channels.  Then, roughly speaking, any channel can be exactly
simulated by a universal channel with the same value of $\xk{\NS}$.
It is this ability of the channels $\mc{U}_{n,m}$ that earns the name  
``universal.''  We need a lemma for this proof:}

\begin{lemma}\label{Tset}
  We call a set $\catchset \subseteq [n]^q$ \emph{$m$-touching} if
  \begin{equation*}
    \forall x_1,\ldots, x_q \in {[n] \choose m} \quad
    \catchset \cap (x_1\times x_2\times\cdots\times x_q) \neq \emptyset.
  \end{equation*}
  There is an $m$-touching set of cardinality $\min \{ n^q, 2nq(\frac{n}{m})^q \}$.
\end{lemma}
\begin{proof}
  If a set is populated by picking $r = 2nq(\frac{n}{m})^q$ elements of $[n]^q$ picked uniformly at random (with replacement), the probability that it is not $m$-touching is bounded above by
  \begin{align*}
    P_{\mt{fail}} \leq \binom{n}{m}^q\left(1-\left(\frac{m}{n}\right)^q\right)^r.
  \end{align*}
  With the simple estimates $\binom{n}{m} \leq 2^n$ and $\ln(1-x) \leq -x$,
  \begin{align*}
    \ln P_{\mt{fail}}
	&\leq q n \ln 2 - r \left(\frac{m}{n}\right)^q\\
	&= (\ln 2 - 2)qn < 0,
  \end{align*}
  so a set with the desired property and cardinality must exist. Indeed, the probability that a set chosen in the way described above isn't $m$-touching is exponentially small in $qn$.
\end{proof}

\begin{proposition}\label{uchanrate}
  For any universal channel $\mc{U}_{n,m}$,
  \begin{equation*}
    \xk{\SR}(\mc{U}_{n,m})
    = \xk{\NS}(\mc{U}_{n,m})
    = \log\frac{n}{m}.
  \end{equation*}
\end{proposition}
\begin{proof}
  By definition, $\xk{\SR}(\mc{U}_{n,m}) \geq \xk{\NS}(\mc{U}_{n,m})$, so it suffices to exhibit a protocol using only shared randomness that achieves this bound. To be precise, for $q$ copies of the channel, we prove the existence of such a protocol which uses the transmission of one of
  \begin{equation*}
    k =\min\left\{
         n^q, \left\lfloor 2qn \left(\frac{n}{m}\right)^q \right\rfloor
       \right\}
  \end{equation*}
  symbols. The rate is $\leq \log\frac{n}{m} + \frac{1}{q} \log 2qn$, which approaches $\log\frac{n}{m}$ as $q\rightarrow\infty$.

The protocol works as follows: Alice and Bob agree on an $m$-touching set $\catchset$ of size $k$ (see Lemma~\ref{Tset}).  They  share randomness in the form of $q$ uniformly random permutations $\pi_1,\ldots,\pi_q \in S_n$. On input $(x_1, \ldots, x_q)$ Alice picks a uniformly random element $(y_1,\ldots ,y_q) \in \catchset^{\pi_1,\ldots,\pi_q} \cap (x_1\times x_2\times\cdots\times x_q)$, where
  \begin{align*}
    \catchset&^{\pi_1,\ldots,\pi_q}\\
	&= \bigl\{ (\pi_1(z_1),\ldots ,\pi_q(z_q)) \ :
              \ \forall\ (z_1,\ldots, z_q)\in \catchset
	\bigr\}
  \end{align*}
  is the set $\catchset$ with its elements permuted according to $\pi_j$ in coordinate $j=1,\ldots,q$. The intersection is guaranteed to exist because $T^{\pi_1,\ldots,\pi_q}$ is also $m$-touching. To send $y$, she only needs a number from $1$ through $k$ to specify the location within $\catchset^{\pi_1,\ldots,\pi_k}$ since the latter is known to Bob.

  This protocol evidently simulates an $S_n^{\times q}$-covariant channel with the property that $\mc{N}(y_1\ldots y_q|x_1\ldots x_q) = 0$ whenever $y_1\ldots y_q \not\in x_1\times x_2\times\cdots\times x_q$. As discussed before, this means that the simulated channel must be $\mc{N}^{\ox q}$.
\end{proof}

\medskip%

\begin{theorem}
\label{srns}
  For any channel $\mc{N} \in \CCs{X}{Y}$,
  \begin{equation*}
    \xk{\SR}(\mc{N}) = \xk{\NS}(\mc{N}) = \log \sum_y \max_x \mc{N}(y|x).
  \end{equation*}
\end{theorem}
\begin{proof}
  First, suppose all the entries of $\mc{N}(y|a)$ are rational numbers, with common denominator $M$, so that $\mc{N}(y|x) = \frac{1}{M} t(y|x)$ for integers $t(y|x)$. Split up each output symbol $y$ into $t_y := \max_x t(y|x)$ many, denoted $(y,j)$, with $j=1,\ldots,t_y$. Now define a new channel by letting $\overline{\mc{N}}\bigl( (y,j)| x \bigr)$ be either $0$ or $1/M$, in such a way that $\mc{N} = \Pi\circ\overline{\mc{N}}$ with the projection map/channel $\Pi:(y,j) \mapsto y$. Clearly $\overline{\mc{N}}$ is a sub-channel (i.e. a restriction on the input alphabet) of the universal channel $\mc{U}_{N, M}$ with $N = \sum_y t_y$. It can therefore be exactly simulated using shared randomness by the protocol of Proposition~\ref{uchanrate}. This requires asymptotic communication rate
  \begin{equation*}
    \log\left( \frac{1}{M}\sum_y t_y \right)
    = \log\left( \sum_y \max_x N(y|x) \right),
  \end{equation*}
  which is precisely the lower bound set by $\xk{\NS}(\mc{N})$, so the claim holds for rational $\mc{N}$.

  For the general case, pick a large integer $M$ and let, for all $x\in X$, $y\in Y$, $t(x|y) := \lfloor M \mc{N}(y|x) \rfloor$. Now adjoin new elements $x'$ ($x\in X$) to the output alphabet, i.e.~define $\widetilde{Y} := Y \cup X'$ and a new channel $\widetilde{\mc{N}}:X \rightarrow \widetilde Y$ with
  \begin{align*}
    \widetilde{\mc{N}}(y|x)  &:= \frac{1}{M}t(y|x), \\
    \widetilde{\mc{N}}(x'|x) &:= 1-\sum_y \widetilde{\mc{N}}(y|x).
  \end{align*}

  Now, $\mc{N}$ can be simulated by $\widetilde{\mc{N}}$ using post-processing by Bob only: if $y\in Y$ is obtained, then it is left alone; if $x'\in X'$ is seen, then Bob uses local randomness to output $y$ with probability
  \begin{equation*}
    Q(y|x')
    = \frac{1}{\widetilde{\mc{N}}(x'|x)}
      \bigl( \mc{N}(y|x)-\widetilde{\mc{N}}(y|x) \bigr).
  \end{equation*}

	So, extending $Q$ to a proper channel by letting $Q(y|y') = 1$ for $y' \in Y$ iff $y = y'$, we have $\mc{N} = Q\circ \widetilde{\mc{N}}$.

  Second, the cost of simulating $\widetilde{\mc{N}}$ is
  \begin{equation*}
    \begin{split}
      &\log\left(
        \sum_{x\in\widetilde Y} \max_x \widetilde{\mc{N}}(x|x)
      \right)\\
      =&\log\left(
        \sum_{y\in Y} \max_x \widetilde{\mc{N}}(x|x)
          + \sum_{x\in X} \widetilde{N}(x'|x)
        \right) \\
      \leq&\log\left(
        \sum_{y\in Y} \max_x \mc{N}(y|x) + \frac{1}{M}|X||Y| \right).
    \end{split}
  \end{equation*} Letting $M \rightarrow \infty$, this rate approaches $\log\left( \sum_y \max_x \mc{N}(y|x) \right) = \xk{\NS}(\mc{N})$.
\end{proof}

To illustrate the idea, if $\mc{N}_0 \prec \mc{N}_1$ denotes the partial order on channels ``$\mc{N}_0$ is equal to $\mc{R}\circ\mc{N}_1\circ\mc{Q}$, for some channels $\mc{R}$, $\mc{Q}$'', the proof uses
  \begin{align*}
    \mc{N} =
	\frac{1}{5}
      \left(
      \begin{array}{cc}
        4 & 2 \\
        1 & 3
      \end{array}
    \right)
    \prec
    \frac{1}{5}
    \left(
      \begin{array}{cc}
        1 & 1 \\
        1 & 1 \\
        1 & 0 \\
        1 & 0 \\
        0 & 1 \\
        0 & 1 \\
        1 & 1
      \end{array}
    \right)
    \prec
    \mc{N}_{7,5}
  \end{align*}
  to show that $\xk{\SR}(\mc{N}) \leq
  \xk{\SR}(\mc{N}_{7,5}) = \log 7/5 =
  \xk{\NS}(\mc{N}) \leq \xk{\SR}(\mc{N})$.

\begin{remark}
	The classical reverse Shannon theorem \cite{CRST} also yields
	an exact simulation of the noisy channel using shared randomness.
	The difference to our result is explained by the different way
	to account for the communication: whereas \cite{CRST} shows that the
	\emph{expected} rate of communication (with respect to the shared
	randomness used in the protocol) is the normal Shannon capacity of the channel, here we consider the much more
	stringent \emph{worst case} communication cost.
\end{remark}

\subsection{Weak simulation and reversibility}
Looking over the formulas for $\zec{\NS}$ and $\xk{\NS}$ of a channel $\mc{N}$, we notice that the former only depends on the channel hypergraph, while the latter actually involves the transition probabilities. Hence it is not surprising that the former is typically strictly smaller than the latter. However if we are content with the simulation of any channel that has the same hypergraph, we recover reversibility:

\begin{proposition}
  Let $\mc{N} \in \CCs{X}{Y}$ with channel hypergraph $H(\mc{N})$ (having hyperedges $\{ e_y : y \in Y \})$. Then,
  \begin{equation*}
    \inf \bigl\{
      \xk{\NS}(\mc{M}) : H(\mc{M}) = H(\mc{N})
    \bigr\}
    = \log \omega^*(H(\mc{N}))
  \end{equation*}
  where $\omega^*(H(\mc{N}))$ is the fractional covering number of the hypergraph of the channel. Since the fractional covering number is equal to the fractional packing number $\alpha^*(H)$, this minimum rate is also equal to $\zec{\NS}(\mc{N})$.
\end{proposition}
\begin{proof}
  Recall the formula for $\xk{\NS}(\mc{N})$: it is the logarithm of the value of the following linear program (all variables understood as non-negative):
  \begin{equation*}
    \min \left\{ \sum_{y\in Y} w(y) : \forall x,y\ w(y) \geq \mc{N}(y|x) \right\}.
  \end{equation*}
  The additional minimisation over channels with prescribed hypergraph $H$ is also a linear program:
  \begin{align*}
    \min \big\{&\sum_{y\in Y} w(y) : w(y) \geq \mc{N}(y|x),\\
	&\mc{N}(y|x)=0 \text{ if } x \notin e_y,\\
	&\sum_y \mc{N}(y|x) = 1 \big\}.
  \end{align*}
  But this is evidently equivalent to
  \begin{equation*}
    \min \left\{ \sum_{y\in Y} w(y) : \forall x \in X, \sum_{y \text{ with } e_y \ni x} w(y) \geq 1 \right\},
  \end{equation*}
  which is exactly the fractional covering number $\omega^*(H)$.
  For the other statements see Proposition~\ref{prop:duality1}.
\end{proof}

\section{Conclusion}
Let us summarise the results (both our own, and others) discussed in this paper. For zero-error communication, we found both the one-shot and asymptotic non-signalling assisted capacities, upper bounding the chains of operationally obvious inequalities:
\begin{align*}
	\alpha(G(\mc{N})) &= \ozec{}(\mc{N}) = \ozec{\SR}(\mc{N})\\
	&\leq \ozec{\SE}(\mc{N}) \leq \ozec{\NS}(\mc{N}) = \lfloor\alpha^\ast(H(\mc{N}))\rfloor
\end{align*}

\begin{align*}
	\log(\Theta(\mc{N})) &= \zec{}(\mc{N}) = \zec{\SR}(\mc{N}) \leq \zec{\SE}(\mc{N})\\
	&\leq \zec{\NS}(\mc{N}) = \log \alpha^\ast(H(\mc{N})) )
\end{align*}

These upper bounds on the entanglement assisted capacities from non-signalling are improved upon by the results of \cite{beigi} and \cite{DSW} which show that the Lov\'{a}sz theta bound applies even in the entanglement assisted case:
\begin{align*}
	\ozec{\SE}(\mc{N}) \leq \lfloor \vartheta(\mc{N}) \rfloor,
	\zec{\SE}(\mc{N}) \leq \log \vartheta(\mc{N}).
\end{align*}

While we proved that $\ozec{}(\mc{N}) \leq \ozec{\SE}(\mc{N})$ can be strict, we don't yet know whether the same can be said of the asymptotic rates, and regard this as one of the main open problems.

In the reverse problem of exactly simulating noisy channels, the non-signalling assisted case was again completely soluble, providing \emph{lower} bounds on the chain
\begin{align*}
	\left\lceil \sum_y \max_x \mc{N}(y|x) \right\rceil
	&= \oxk{\NS}(\mc{N}) \leq \oxk{\SE}(\mc{N})\\
	& \leq \oxk{\SR}(\mc{N}) \leq \oxk{\NC}(\mc{N}).
\end{align*}

For each inequality in this chain of one-shot costs, a channel showing that it can be strict was exhibited. Some open questions remain regarding the potential sizes of these separations (see section~\ref{sec:sim}).

For the asymptotic rates of communication things were shown to be simpler: While large gaps can exist between the costs with free shared randomness and without ($\xk{\NC}(\mc{N}) \geq \log \rank \mc{N}$), given free correlations from \emph{any} class of non-signalling correlations which contains shared randomness the rates are equal:

\begin{align*}
	\log \left(\sum_y \max_x \mc{N}(y|x) \right)
	&= \xk{\NS}(\mc{N}) = \xk{\SE}(\mc{N})\\
	&= \xk{\SR}(\mc{N}) \leq \xk{\NC}(\mc{N}).
\end{align*}

%
%
\section*{Acknowledgments}
We would like to thank
Nicolas Brunner,
Runyao Duan,
Tsuyoshi Ito,
Ashley Montanaro,
Marcin Paw\l{}owski,
Paul Skrzypczyk and
Stephanie Wehner
for useful discussions.
%


\appendix[Notations]
\noindent $[n]$: The set $\{1,\ldots,n\}$.

\noindent $\CCs{X}{Y}$: The set of classical channels with input alphabet $X$ and output alphabet $Y$.

\noindent $\aBCC{\mathbf{C}}{A}{S}{B}{T}$: The set of bipartite classical channels with input alphabets $A$ (for Alice) and $B$ (for Bob) and respective output alphabets $S$ and $T$.

\noindent $\arbclass$: Some class of correlations: one of $\NS = $ non-signalling, $\SE =$ shared entanglement, $\SR=$ shared randomness, $\NC $ (or ommited) = no correlation.

\noindent $\aBCC{\arbclass}{A}{S}{B}{T}$: The subset of $\aBCC{\mathbf{C}}{A}{S}{B}{T}$ in the class $\arbclass$.

\noindent $\mc{N}(y|x)$: The probability that the channel $\mc{N}$ outputs symbol $y$ when symbol $x$ is input.

\noindent $E(G)$: Edges of the graph $G$.

\noindent $E(H)$: Hyperedges of the hypergraph $H$.

\noindent $G(\mc{N})$: Confusability graph of the channel $\mc{N}$.

\noindent $H(\mc{N})$: Hypergraph of the channel $\mc{N}$.

\noindent $\chi(G)$: The clique hypergraph of the graph $G$.

\noindent $\alpha(G)$: The independence number of the graph $G$.

\noindent $\alpha^*(H)$: The fractional packing number of the hypergraph $H$.

\noindent $\omega^*(H)$: The fractional covering number of the hypergraph $H$.

\noindent $\ozec{\arbclass}(\mc{N})$: One-shot zero-error capacity of $\mc{N}$ assisted by $\arbclass$.

\noindent $\zec{\arbclass}(\mc{N})$: Zero-error capacity of $\mc{N}$ assisted by $\arbclass$.

\noindent $\oxk{\arbclass}(\mc{N})$: One-shot simulation cost of $\mc{N}$ assisted by $\arbclass$.

\noindent $\xk{\arbclass}(\mc{N})$: Simulation cost of $\mc{N}$ assisted by $\arbclass$.

\appendix[Pair-wise versus mutual distinguishability for sets of local residual states of correlations]
\label{nscomb}

\begin{definition}
  We say that two classical channels $\mc{N}$ and $\mc{M}$ in $\CCs{X}{Y}$ are \emph{pair-wise distinguishable}, and write $\mc{N} \bowtie \mc{M}$ if there is an input $x^{\ast} \in X$ such that
  \begin{equation*}
    \sum_{y\in Y} \mc{N}(y|x^{\ast})\mc{M}(y|x^{\ast}) = 0.
  \end{equation*}
\end{definition}

If Alice makes an input $a$ to her side of a bipartite correlation $P \in \aBCC{\mathbf{C}}{A}{X}{B}{Y}$, and obtains the output $x$ then the conditional distribution on Bob's side is a classical channel $P_{ax}(y|b) \in \CCs{B}{Y}$ where $P_{ax}(y|b)$ is simply the conditional distribution $P(y|b,a,x)$ given by Bayes rules, but written differently to emphasise the fact that we regard $a$ and $x$ as fixed. Similarly, there are $|B||Y|$ such conditional channels $P_{by}(x|a)$ on Alice's side.

\begin{definition}
  We say that a class of bipartite correlations $\arbclass$ has property $\pw_A$ if the existence of a correlation $P(x,y|a,b) \in \XBs{A}{X}{B}{Y}$, and $S \subseteq A \times X$, satisfying
  \begin{equation*}
    P_{ax} \bowtie P_{a'x'} \quad \forall (a,x), (a',x') \in S
  \end{equation*}
  implies the existence of another correlation, $P'(x,y|a,b) \in \XBs{A}{X}{B\cup \{b^{\ast}\}}{Y\cup Y'}$, which is identical to $P$ when restricted to the input alphabets of $P$,
  \begin{align*}
    \forall a &\in A, x \in X, b \in B, y \in Y:\\
    &P'(x,y|a,b) = P(x,y|a,b),
  \end{align*}
  but which has an extra input $b^{\ast}$ on Bob's side such that
  \begin{align*}
    \forall &(a,x), (a',x') \in S :\\
    &\sum_{y \in Y\cup Y'} P'_{ax}(y|b^{\ast})P'_{a'x'}(y|b^{\ast}) = 0.
  \end{align*}
  $\arbclass$ has property $\pw_B$ if it satisfies the same condition with the roles of the parties reversed. If $\arbclass$ has property $\pw_A$ and property $\pw_B$ then we simply say it has property $\pw$.
\end{definition}
In other words, if a correlation $P$ belongs to a $\pw_A$ class $\arbclass$, and the graph induced on $A\times X$ by the pair-wise distinguishability relation between the conditional states $P_{ax}$ associated with the vertices contains a clique $S$, then there is another correlation $P'$ in $\arbclass$ which behaves like $P$ except that Bob has some extra output symbols (possibly), and one new input symbol $b^{\ast}$, which when input, yields pair-wise orthogonal output distributions on $Y$ for all elements of $S$ so that they can be perfectly distinguished simultaneously.

To illustrate this idea, we now show that the class $\NS$ of generalised non-signalling correlations is not $\pw$: If Alice and Bob's shared correlation $P$ is the PR-box, then the conditional channels $P_{ax}$ are given by
\begin{equation*}
  P_{ax}(y|b) = [x\oplus y = (a \wedge b)].
\end{equation*}
These channels are all pair-wise distinguishable, but of course, the required input on Bob's side depends on the pair. If NS were $\pw$, then the existence of $P \in \NS$ would imply the existence of another correlation in $\NS$ where a single input on Bob's side would suffice to distinguish the $4$ residual states. But obviously this would allow Bob to determine Alice's input, so this is a contradiction: Put another way, if a class is $\pw$ and contains the PR-box then it also contains signalling correlations.

On the other hand,
\begin{proposition}
  The class of bipartite correlations which can be implemented as local measurements on entangled quantum states ($\SE$) is $\pw$.
\end{proposition}
\begin{proof}
  Assuming w.l.o.g. that Alice measures first: Alice inputs $a$ (corresponding to her measuring of some POVM on her side) and obtains outcome $x$, leaving a residual state $\rho_{ax}$ on Bob's side. The conditional channel $P_{ax}(y|b)$ is given by
  \begin{equation*}
    P_{ax}(y|b) = \tr \rho_{ax} B^{(b)}_y,
  \end{equation*}
  so if $P_{ax} \bowtie P_{a'x'}$ then there must be some input $b$ on Bob's side, corresponding to a POVM with elements $\{B_y\}_{y\in Y}$ say, such that
  \begin{multline*}
    \forall y : P_{ax}(y|b)P_{a'x'}(y|b)\\
    = \left( \tr B_y \rho_{ax}\right) \left(\tr B_y \rho_{a'x'} \right)
    = 0
  \end{multline*}
  which implies that the residual states $\rho_{ax}$ and $\rho_{a'x'}$ are orthogonal (i.e. have disjoint support).

  A clique of pair-wise distinguishable conditional channels on Bob's side therefore corresponds to a clique of \emph{mutually} orthogonal residual states on his side. Therefore, there is a single measurement which perfectly distinguishes all members of the clique, which we can obviously use to construct a correlation $P'$ in the class of correlations $\SE$ with the required properties.
\end{proof}

From this result and the previous example, it is clear that the PR-box cannot be perfectly implemented by shared entanglement (a fact which can alternatively be proved by the Tsirelson bound).

\begin{proposition}\label{pw-G}
	If $\arbclass$ is a PW class of correlations, then the one-shot (and hence asymptotic) $\arbclass$-assisted zero-error capacities $\ozec{\arbclass}(\mc{N})$ and $\zec{\arbclass}(\mc{N})$ of a channel $\mc{N}$ only depend on the confusability graph $G(\mc{N})$.
\end{proposition}
\begin{proof}
	Let $P$ be a correlation in $\aBCC{\arbclass}{[c]}{X}{Y}{[c]}$ such that the standard `wiring' yields the largest possible identity channel i.e.\ one with $c = \ozec{\arbclass}(\mc{N})$ symbols.
	\[
		\sum_{x,y}P(x,\hat{z}|z,y) =
		\begin{cases}
			1 \text{ if } z = \hat{z}
			\\ 0 \text{ otherwise. }
		\end{cases}
	\]

	Write $\delta_y (x,x')$ if there is a single input $y$ on Bob's side such that $\forall z\neq z', \forall x: P_{zx:y} \perp P_{z'x:y}$: In words, $\delta_y (x,x')$ means that if Bob knows that the channel input was one of $x$ or $x'$ then he can distinguish which $z$ Alice chose by making input $y$ to his side of the correlation.

	When Bob gets output symbol $y$ from the channel, let $e_y$ denote the set of possible inputs. He can decode $z$ perfectly iff $\delta_y(x,x') \forall x \neq x' \in e_y$. If we draw a graph on $X$ with edges labelled by outputs $Y$, with a $y$-edge between $x$ and $x'$ iff $\delta_y(x,x')$, then (ignoring edge labels and multiplicities) this graph must contain $G(\mc{N})$.

	Recalling the discussion after Definition \ref{cg}, we know that $\ozec{\arbclass}(\chi(G(\mc{N}))) \leq \ozec{\arbclass}(H(\mc{N}))$. By the PW property of $\arbclass$ it must be possible to find a new correlation in $\arbclass$ such that if Bob knows that $x$ was in \emph{any} clique in this graph, then he can still determine $z$. So the $\arbclass$-assisted zero-error capacity of the clique hypergraph of $G$ is at least $c$.

	Therefore, if $\arbclass$ is PW then
	\[
		\ozec{\arbclass}(\chi(G(\mc{N}))) = \ozec{\arbclass}(H(\mc{N})),
	\]
	so the zero-error capacity depends only on the confusability graph.
\end{proof}

\begin{proposition}\label{pw-bound}
  For a correlation $P \in \aBCC{\mathbf{C}}{A}{X}{B}{Y}$, let $\Delta_B$ ($\Delta_A$) be the graph on the $|A||X|$ ($|B||Y|$) conditional channels on Bob's (Alice's) side where edges denote pair-wise distinguishability. If $P$ belongs to a class which is both $\pw$ and non-signalling, then
  \begin{equation*}
    \bar{\chi}(\Delta_B) \geq |A|
  \end{equation*}
  and
  \begin{equation*}
    \bar{\chi}(\Delta_A) \geq |B|
  \end{equation*}
  where $\bar{\chi}$ denotes the clique covering number. In particular, these bounds apply to the correlation class $\SE$: The set of bipartite correlations which can be implemented using entanglement.
\end{proposition}
\begin{proof}
  Let $A^{\ast}$ ($B^{\ast}$) be a minimal clique covering of $\Delta_A$ ($\Delta_B$). Suppose that $P$ is in $\arbclass$ which is $\pw$ and non-signalling. By repeated use of the definition of the $\pw$ property, $\arbclass$ contains a $P \in \aBCC{\mathbf{C}}{A\cup A^{\ast}}{X \cup X'}{B\cup B^{\ast}}{Y \cup Y'}$ such that for $q \in B^{\ast}$
  \begin{align*}
    \forall y \in Y \cup Y', (a,x) \in q, (a',x') \in q :\\
    P'_{ax}(y|q)P'_{a'x'}(y|q) = 0.
  \end{align*}
  This means that if Alice inputs $a \in A$ to $P'$ and obtains output $x$, and then tell's Bob a clique in $B^{\ast}$ which contains $(a,x)$, Bob can determine $(a,x)$ exactly. In particular, he discovers Alice's choice of input in $A$ without error, by Alice transmitting one of $|B^{\ast}| = \bar{\chi}(\Delta_B)$ messages. Since non-signalling correlations can't increase the zero-error capacity of identity channels (a simple consequence of Theorem~\ref{NSZELP}), if $\bar{\chi}(\Delta_B) < |A|$ then $P'$ cannot be non-signalling (and similarly if $\bar{\chi}(\Delta_A) < |B|$).
\end{proof}

\end{document}